\documentclass[
  aps,
  physrev,
  reprint,
  superscriptaddress,
  nofootinbib,
  amsmath,
  amssymb
]{revtex4-2}

\usepackage[normalem]{ulem}
\usepackage{graphicx}
\usepackage{bm}
\usepackage{mathrsfs}
\usepackage[caption=false]{subfig}
\usepackage{hyperref}
\usepackage{orcidlink}

\newcommand{\review}[1]{\textcolor{blue}{#1}}
\renewcommand{\review}[1]{#1} % Uncomment this line for final version
\newcommand{\remove}[1]{\textcolor{red}{\sout{#1}}}
\renewcommand{\remove}[1]{} % Uncomment this line for final version

\begin{document}

% Full title of the paper (Capitalized)
\title{UPLOAD-HELIX: A High-Helicity Single-Mode Microwave Haloscope with Low-Noise Interferometric Readout for Ultralight Axion Dark Matter}

\author{Robert C. Crew\,\orcidlink{0009-0002-2143-3422}}
\affiliation{%
  Quantum Technologies and Dark Matter Research Laboratory, The University of Western Australia, Perth, WA 6009, Australia
}

\author{Emma C.I. Paterson\,\orcidlink{0000-0003-3752-5815}}
\affiliation{%
  Quantum Technologies and Dark Matter Research Laboratory, School of Physics, The University of Western Australia, Perth, WA 6009, Australia
}

\author{Maxim Goryachev\,\orcidlink{0000-0002-0257-4054}}
\affiliation{%
  Quantum Technologies and Dark Matter Research Laboratory, The University of Western Australia, Perth, WA 6009, Australia
}

\author{Eugene N. Ivanov\,\orcidlink{0009-0006-7712-9330}}
\affiliation{%
  Quantum Technologies and Dark Matter Research Laboratory, The University of Western Australia, Perth, WA 6009, Australia
}

\author{Pashupati Dhakal\,\orcidlink{0000-0002-9381-4091}}
\affiliation{%
  Thomas Jefferson National Accelerator Facility, Newport News, USA
}

\author{Tugrul Talha Ersoz\,\orcidlink{0000-0002-0365-6756}}
\affiliation{%
  University of Birmingham, School of Metallurgy and Materials, Edgbaston, Birmingham B15 2TT, United Kingdom
}
\affiliation{%
  Department of Metallurgical and Materials Engineering, Selçuk University, 42075, Konya, Turkey
}

\author{Michael E. Tobar\,\orcidlink{0000-0002-3139-1994}}
\email{michael.tobar@uwa.edu.au}
\affiliation{%
  Quantum Technologies and Dark Matter Research Laboratory, The University of Western Australia, Perth, WA 6009, Australia
}

\author{Jeremy F. Bourhill\,\orcidlink{0000-0002-5667-7745}}
\affiliation{%
  Quantum Technologies and Dark Matter Research Laboratory, The University of Western Australia, Perth, WA 6009, Australia
}

\begin{abstract}
  We propose a superconducting single-mode microwave haloscope based on \remove{helical}\review{chiral} cavity resonators for the detection of ultralight dark matter axions over the mass range \remove{$10^{-18}$--$10^{-13}\,\mathrm{eV}$}\review{$4\times10^{-19}$--$~4\times10^{-14}\,\mathrm{eV}$}. \review{Building on the single-mode chiral-cavity concept introduced by Bourhill \textit{et al.}~\cite{Bourhill2023}, we develop a resonator geometry compatible with subtractive manufacturing from high-purity bulk niobium, taking advantage of the substantially lower surface resistance achievable relative to the additively manufactured Möbius cavity proposed in the earlier work. An inverse-design framework is then used to maximise a figure of merit derived to minimise the measurement time required to achieve a fixed experimental sensitivity. The resulting optimised bulk-niobium design achieves a figure of merit more than three orders of magnitude larger than the additively manufactured Möbius benchmark.} \remove{Building on the single-mode helical-cavity concept introduced by Bourhill \textit{et al.}~\cite{Bourhill2023}, we use an inverse-design framework to develop practical resonator geometries compatible with superconducting niobium fabrication. The optimisation employs a figure of merit derived to minimise the measurement time required to achieve a fixed experimental sensitivity. Relative to the heuristic Möbius-geometry benchmark, the best subtractively manufacturable bulk-niobium design achieves a figure of merit more than three orders of magnitude larger.} An experimentally informed microwave interferometric readout model, incorporating measured electronics noise and active suppression of pump amplitude noise, is used to project the sensitivity of the proposed experiment. For an acquisition time of three months, the haloscope is projected to reach $g_{a\gamma\gamma}<10^{-11}\,\mathrm{GeV}^{-1}$ across more than four orders of magnitude in axion mass. The projected sensitivity extends approximately one order of magnitude below the current exclusion limits set by CAST, providing a practical pathway towards a high-sensitivity direct search for ultralight dark matter axions.
\end{abstract}


\maketitle

\section{Introduction}

\subsection{Searching for Ultralight Dark Matter Axions with Resonant Cavities}
\remove{Among the leading dark matter candidates are ultralight dark matter (ULDM) axions, which}\review{QCD axions arise from the Peccei-Quinn solution}\remove{ have been proposed as a solution} to the strong charge-parity (CP) problem in quantum chromodynamics (QCD)~\cite{Peccei1977,PQ1977b,Weinberg1978,Wilczek1978}\remove{ and}\review{. Generally axion-like particles (ALPs) need not solve the strong CP problem but} may \review{likewise} account for the entire dark matter content of the Universe~\cite{Di-Luzio:2021wu,Di_Luzio_2021,Sokolov:2021uv,Visinelli19}. Axions can couple to photons through the axion-photon chiral anomaly, \review{which in SI units can be }described by the interaction term \remove{$g_{a\gamma\gamma}a~\mu_0\,\mathbf{E}\cdot\mathbf{H}$}\review{in the Lagrangian density $\mathcal{L}_{a\gamma\gamma} = \epsilon_0 c g_{a\gamma\gamma} a\mathbf{E}\cdot\mathbf{B}$}~\cite{Peccei2008}. This interaction allows an oscillating axion field, $a$, to induce measurable electromagnetic signals $(\mathbf{E},\mathbf{B})$ in \remove{suitably designed resonant}\review{suitable} structures, through the axion-photon coupling constant, $g_{a\gamma\gamma}$. \review{In a traditional static-field cavity-based axion haloscope~\cite{Sikivie1985} a strong static magnetic field is employed to induce the conversion of axions into microwave photons within a resonant microwave cavity, and the sensitivity is parametrised by the form factor quantifying the overlap between the electric field of the cavity and the external magnetic field.}\remove{The strength of the resulting signal is determined by a dimensionless form factor, which quantifies the spatial overlap between the electromagnetic mode of the resonator and the effective axion-induced source field. Consequently, numerous experiments have been developed to search for axion dark matter using resonant electromagnetic systems.} \review{ Beginning in the late 1980s~\cite{depanfilis_limits_1987, hagmann_results_1990} numerous implementations of the static field microwave cavity haloscope have been developed over the approximately $1$ to $100~\mu\mathrm{eV}$ mass range. Representative examples include ADMX~\cite{braine_extended_2020, du_search_2018}, CAPP~\cite{ahn_extensive_2024,kim_near-quantum-noise_2023}, HAYSTAC~\cite{backes_quantum_2021, bai_dark_2025}, and ORGAN~\cite{quiskamp_direct_2022, quiskamp_near-quantum-limited_2025}.}

\remove{Traditional cavity-based axion searches employ a strong static magnetic field to induce the conversion of axions into microwave photons within a resonant cavity, and the sensitivity is parametrised by the form factor quantifying the overlap between the electric field of the cavity and the external DC magnetic field.} While this approach has achieved exceptional sensitivity\review{~\cite{goodman_admx_2025}}, the requirement for large \remove{superconducting magnets}\review{external magnetic fields} \remove{precludes}\review{limits} the use of low-loss superconducting resonators~\cite{Creedon16,Romanenko20,Posen2020,McA21}, thereby limiting the achievable cavity quality factor, $Q$, and ultimately the experimental sensitivity. \review{Although type-II superconductors, which can maintain superconductivity in the presence of high magnetic fields, have been demonstrated~\cite{posen_high-quality-factor_2023,ahyoune_rades_2025,ahn_biaxially_2022} their achievable quality factors under such fields are orders of magnitude below the best niobium resonators in a low field environment~\cite{Romanenko20}.}

\review{The key limitation of the conventional static-field empty microwave cavity haloscope is that }\remove{T}\review{t}he cavity must \remove{also operate}\review{be resonant} at the axion\review{-Compton frequency.} \remove{frequency precluding searches of ULDM that would require unreasonably large cavities.} \review{As such, for sufficiently low-mass axions a conventional cavity would need impractically large dimensions, motivating alternative techniques for ultralight dark matter (ULDM) axions. Nonetheless ULDM axions are well motivated~\cite{Di-Luzio:2021wu}. Searches for ULDM axions have employed broadband magnetometry~\cite{arza_earth_2022,friel_search_2024, nishizawa_axion_2026}. Resonant systems such as the DANCE experiment employ an optical bow-tie cavity~\cite{obata_optical_2018,oshima_first_2023}, and the DMRadio program employs lumped LC resonators~\cite{rapidis_status_2023}. Helioscopes~\cite{altenmuller_new_2024} and astrophysical observations~\cite{sisk-reynes_new_2021, dessert_no_2022} give complementary constraints, although these depend on the stellar, plasma, or magnetic field models used~\cite{matthews_how_2022}.}

Alternative approaches \review{to push microwave cavity-based axion searches to lower masses} have been proposed that exploit axion-mediated coupling between resonant electromagnetic modes of non-zero overlap integral~\cite{sikivie_superconducting_2013, McAllister2016, goryachev_axion_2019}. Such upconversion experiments are sensitive to axion frequencies comparable to the frequency difference between the modes, and have been demonstrated to be sensitive to axions as either frequency modulation~\cite{UPLOAD2021} or additional power~\cite{Thomson2023} in the readout mode. These two detection approaches have been used to constrain axion masses over the ranges $7.44$–$19.38~\mathrm{neV}$~\cite{UPLOAD2021} and $1.12$–$1.20~\mu\mathrm{eV}$~\cite{Thomson2023}, respectively. However, practical implementations of two-mode upconversion schemes are constrained by feedthrough, mode hybridisation, nonlinear effects, and injection locking, which limit the minimum achievable mode detuning and consequently prevents sensitivity to the lowest axion masses~\cite{Bourhill2023,Thomson2023}.

A possible solution to this limitation was proposed through the development of twisted cavity resonators~\cite{Bourhill2023}. These resonators support monochromatic modes with nonzero electromagnetic helicity, $\mathscr H$, \remove{a quantity proportional to the degree of orthogonality between spatial components of the $\mathbf{E}$ and $\mathbf{H}$ fields,}\review{defined by the normalised electromagnetic field overlap in~\eqref{eq:helicity_density} and~\eqref{eq:helicity}}\remove{ and whose squared magnitude}\review{. The square magnitude of which} acts as the effective axion-photon form factor\review{~\cite{Bourhill2023}}, allowing the axion field to couple directly to a single resonant eigenmode. Consequently, the resonant mode functions simultaneously as both the pump and signal mode in an upconversion detection scheme. Using axion-modified electrodynamics~\cite{wilczek_two_1987,tobar_modified_2019,tobar_poynting_2022} it can be shown~\cite{Bourhill2023} that the axion signal produces amplitude modulation (AM) sidebands within the resonator bandwidth, $\Delta f$, enabling a single-mode search for ultralight axions~\cite{Bourhill2023} \remove{from $10^{-18}$--$10^{-13.7}~\mathrm{eV}$}\review{over four orders of magnitude in mass from $4\times10^{-19}$-- $4\times10^{-15}~\mathrm{eV}$}. That is, while the cavity may operate in the X-band, \remove{the cavity}\review{it} is broadband sensitive to ultralight axions \review{inducing modulation sidebands about the fixed resonant carrier}. \review{Axion induced sidebands are resonantly enhanced within the loaded cavity bandwidth while the response rolls off outside the bandwidth.} 

\subsection{Optimising the Design of the Twisted \remove{Anyon} \review{Chiral} Cavity Resonator}
The \remove{first }twisted cavity resonators \review{developed in~\cite{Bourhill2023}} were \remove{developed by}\review{constructed from} uniformly twisting prismatic resonators with cross-sections possessing dihedral ($D_n$) symmetry. Twisting the resonator breaks the mirror symmetry of the conducting boundary conditions, inducing magnetoelectric coupling between near-degenerate transverse-electric (TE) and transverse-magnetic (TM) modes~\cite{Bourhill2023,paterson2025electromagnetic,paterson2025dynamicallytuneablehelicitytwisted,paterson_Berry}. This hybridisation gives rise to electromagnetic eigenmodes with nonzero $\mathscr H$. To further reduce microwave losses, the end faces of the twisted resonator were joined to form a M\"obius resonator, eliminating end-boundary losses, reducing one source of dissipation and enabling higher-$Q_0$ helical modes~\cite{Bourhill2023}.

Initial prototypes were aluminium cavities with a $D_3$ cross-section, manufactured using laser powder bed fusion (LPBF). These devices experimentally verified the predicted mode structure and mode-frequency tuning with twist angle, validating the existence of electromagnetic modes with large $\mathscr H$~\cite{Bourhill2023}. While providing an important proof-of-principle demonstration of the underlying physics, \review{preliminary unpublished cryogenic measurements of these cavities yieled intrinsic quality factors} $Q_0$ \remove{values}\review{that} didn't exceed $10^5$ for high helicity modes, even \review{cooled} below the superconducting transition temperature. This motivated investigations into alternative designs to construct higher-$Q$ \remove{helical}\review{chiral} resonators.

Because $\mathscr H$ depends on the three-dimensional spatial overlap of the electric and magnetic field distributions, even small changes to the cavity boundary can produce large changes in the resulting helicity, making it difficult to predict which geometries will support highly helical modes. In this work, we therefore formulate the cavity-design problem as a boundary-shape optimisation problem. Building on the inverse-design framework introduced in Ref.~\cite{Paterson2026InverseDesign}, we optimise candidate cavity geometries directly for ultralight axion detection by maximising a composite figure of merit, FoM~\eqref{eq:FoM}.

The original work further noted that implementing twisted resonators as high-purity niobium superconducting radio-frequency (SRF) cavities could dramatically reduce microwave surface losses, enabling substantially higher $Q_0$, narrower $\Delta f$, and consequently enhanced sensitivity to ultralight axion dark matter~\cite{Bourhill2023}. However, the prototype resonators were intended primarily as proof-of-principle demonstrations rather than high-$Q$ SRF cavity designs. The cavity geometries investigated in this study were therefore developed to be compatible with current routes for fabricating superconducting niobium resonators. Accordingly, the optimisation incorporates practical design constraints associated with the intended manufacturing method, together with dimensional constraints required for operation in a dilution refrigerator at millikelvin temperatures. The resulting cavity designs are then further evaluated with respect to practical considerations, including seam-induced losses.

In this work, we present results from optimised \remove{helical}\review{chiral} cavity resonator designs \remove{specifically }developed for fabrication from superconducting niobium\review{, propose an accompanying interferometric readout architecture for detecting ULDM, and provide projected sensitivities}. Using inverse-design techniques, the cavity geometries are obtained by maximising a\review{n axion-specific} FoM directly proportional to the \review{electronics noise limited SNR within the resonator bandwidth. Under the adopted surface resistance assumptions the bulk niobium design based on subtractive manufacturing provides the largest projected FoM. Our projected limits on axion-photon coupling motivate a direct search for ULDM axions over the mass range examined below.} \remove{SNR for dark-matter axion detection}. \remove{The electromagnetic performance and practical realisability of the resulting designs are investigated, providing a pathway towards high-$Q$ \remove{helical}\review{chiral} resonators for future ULDM searches. To assess the sensitivity of these designs to ULDM we propose a prototype experimental an interferometric readout architecture and provide potential limits on $g_{a\gamma\gamma}$ achievable with these designs.}

\section{Cavity Inverse Design and Optimisation Methodology}

To identify cavity geometries optimised for single-mode ULDM axion searches using resonant electromagnetic modes with nonzero $\mathscr H$, we employ the inverse-design framework of Ref.~\cite{Paterson2026InverseDesign}. In this framework, a PyGAD-based~\cite{gad2023pygad} genetic algorithm (GA) explores a parameterised design space, while COMSOL Multiphysics\textsuperscript{\textregistered}~\cite{comsol_multiphysics_v6.4} finite-element method (FEM) eigenmode simulations are performed via the MPh interface for fitness evaluation. The principal modification to this framework is the fitness function, which employs a composite FoM designed to minimise the measurement time required to achieve a fixed experimental sensitivity in a single-mode ULDM axion search. The following subsections define the key electromagnetic properties of the resonant modes, derive the axion-induced modulation index in terms of these quantities, and show how this naturally leads to the optimisation FoM adopted throughout this work. The optimisation framework is then applied to the candidate cavity geometries discussed in Section~\ref{sec:cavity_geometries}.

\subsection{Derivation of the Optimisation Figure of Merit}
\label{sec:FOMderivation}

\subsubsection{Electromagnetic Helicity}
\label{sec:helicity}

For a monochromatic resonant mode $i$, the local time-averaged helicity density, $h_i(\mathbf r)$, is defined as~\cite{paterson2025electromagnetic,paterson2025dynamicallytuneablehelicitytwisted}
\begin{equation} \label{eq:helicity_density}
        h_i(\mathbf{r}) = \frac{2}{V}\,\mathrm{Im}\!\left[\mathbf{e}_i(\mathbf{r}) \cdot \mathbf{h}_i^{*}(\mathbf{r})\right] = \frac{2\,\mathrm{Im}\!\left[\mathbf{E}_i(\mathbf{r}) \cdot \mathbf{H}_i^{*}(\mathbf{r})\right]}{V\sqrt{\mathcal{E}_i\,\mathcal{M}_i}},
\end{equation}
where $\mathbf{E}_i(\mathbf{r})$ and $\mathbf{H}_i(\mathbf{r})$ are the electric and magnetic vector fields of the mode, and
\begin{equation}
        \mathcal{E}_i=\frac{1}{V}\int \mathbf{E}_i(\mathbf{r})^{*}\cdot\mathbf{E}_i(\mathbf{r})\,dV,
        \qquad
        \mathcal{M}_i=\frac{1}{V}\int \mathbf{H}_i(\mathbf{r})^{*}\cdot\mathbf{H}_i(\mathbf{r})\,dV
\end{equation}
are the volume-averaged squared electric and magnetic field amplitudes over the cavity volume ($V$). The corresponding root-mean-square-normalised eigenvectors are 
\begin{equation}
        \mathbf{e}_i(\mathbf{r}) = \frac{\mathbf{E}_i(\mathbf{r})}{\sqrt{\mathcal{E}_i}}, \qquad
        \mathbf{h}_i(\mathbf{r}) = \frac{\mathbf{H}_i(\mathbf{r})}{\sqrt{\mathcal{M}_i}}.
\end{equation}

The total electromagnetic helicity of the mode, $\mathscr{H}_i$, is then defined as
\begin{equation}\label{eq:helicity}
        \mathscr{H}_i=\int h_i(\mathbf r)\,dV.
\end{equation}
For each eigenmode of a candidate cavity, $\mathscr{H}_i$ can be evaluated directly from the COMSOL eigenmode solution and used by the optimisation algorithm.

\subsubsection{Resonator Bandwidth}
\label{sec:bandwidth}

The intrinsic bandwidth of the resonant mode $i$ is 
\begin{equation}
    \Delta f_i=\frac{f_i}{Q_0^i},
\end{equation}
where $f_i$ and $Q_0^i$ are the resonant frequency and intrinsic quality factor of the mode, respectively. $Q_0^i$ depends on both the cavity geometry and the microwave surface resistance, $R_s$. To separate the influence of the cavity geometry from that of the cavity-wall material properties, we introduce the geometric factor, $G_i$, which depends solely on the electromagnetic field distribution of the resonant mode. For resonant mode $i$, the geometric factor is~\cite{Creedon16}
\begin{equation}
    G_i=
    2\pi f_i
    \frac{\displaystyle \iiint \mu_0\left|\mathbf{H}_i\right|^2\,dV}
         {\displaystyle \iint \left|\mathbf{H}_i^{\tau}\right|^2\,dS},
    \label{eq:geom_factor}
\end{equation}
where $\mu_0$ is the permeability of free space and $\mathbf{H}_i^{\tau}$ is the component of the magnetic field tangential to the cavity surface. $Q_0^i$ is then given by

\begin{equation}
    Q_0^i=\frac{G_i}{R_s}.
    \label{eq:Q_factor}
\end{equation}

The cavity resonators considered in this work are intended for operation in a dilution-refrigerator-based ultralight axion search, where millikelvin temperatures ($<25~\mathrm{mK}$) suppress thermal noise and measurements are performed at low input powers ($<30~\mathrm{dBm}$) to avoid perturbing the weak axion-induced signal. Under these operating conditions, the microwave surface resistance, $R_s$, of superconducting radio-frequency (SRF) cavities fabricated from high-purity niobium, defined here as niobium with a residual resistivity ratio (RRR) greater than 200, is expected to exhibit only a weak frequency dependence over the 1–20$~\mathrm{GHz}$ range considered, subject to the idealised conditions discussed in Appendix~\ref{app:freq_dep_Rs}. Accordingly, the optimisation algorithm treats $R_s$ as frequency independent. However, its magnitude depends on the material and fabrication method used to produce the cavity. For each fabrication method, $\hat{R}_s$ is therefore taken as the lowest microwave $R_s$ reported in the literature for that method. The adopted values of $\hat{R}_s$, together with their justification and underlying assumptions, are discussed in Appendix~\ref{app:Rs_analysis}. These values are used by the optimisation algorithm to estimate the intrinsic quality factor, $\hat{Q}_0^i$, and corresponding bandwidth, $\hat{\Delta f}_i$, of each candidate mode directly from the COMSOL eigenmode solution.

\subsubsection{Axion-Induced Modulation Index}
\label{sec:Axion_Mod_Index}

Having defined the key electromagnetic properties of the resonant mode, we now derive an expression for the axion-induced amplitude modulation index in terms of these quantities. The amplitude modulation index, $m_r$, is given by~\cite{Bourhill2023}
\begin{equation} \label{eq:mam}
    |m_{r}| = \frac{1}{2\sqrt{2}} Q_L\frac{\omega_a}{\omega_0} \left<\theta_0\right> |\mathscr{H}|
\end{equation}
where $Q_L$ is the loaded quality factor of the cavity given by $Q_L = Q_0/(\beta+1)$, with $\beta$ the coupling coefficient. $\omega_a$ is the axion angular frequency, and $\omega_p$ is the angular frequency of the applied pump. Throughout this work, the pump is assumed to be resonant with the cavity mode, such that $\omega_p=\omega_0$, where $\omega_0$ is the resonant angular frequency of the high helicity mode. The root-mean-square amplitude of the axion field is given by
\begin{equation} \label{eq:thetaUnit}
    \left<\theta_0\right>^2 = g_{a\gamma\gamma}^2\rho_a \frac{c^3 \hbar}{\omega_a^2}
\end{equation}
where $\rho_a$ is the axion density which we take to be $4.5\times10^5~\mathrm{GeV/m^3}$, $c$ is the speed of light, and $\hbar$ is the reduced Planck's Constant used here in $\mathrm{GeVs}$. Note that haloscopes operate in the limit where the wavelength of the axion field is much larger than the cavity size, hence the axion field is taken to be $\Theta (t) = \sqrt{2} \left<\theta_0\right> \cos(\omega_a t)$. Employing~\eqref{eq:thetaUnit} and noting that~\eqref{eq:mam} is only valid inside the resonator bandwidth, so we must replace $Q_L$ with the general $Q_L/\sqrt{1 + 4 Q_L^2 (\omega_a/\omega_p)^2}$ we write
\begin{equation}
    |m_{r}| = \frac{\sqrt{\rho_a c^3 \hbar}}{2\sqrt{2}} g_{a\gamma\gamma}\cdot\frac{Q_L}{\sqrt{1 + 4 Q_L^2(\frac{\omega_a}{\omega_p})^2}}\frac{|\mathscr{H}|}{\omega_0}.
\end{equation}
To simplify later analysis we recast this in terms of resonator frequency $f_0 = \omega_0/(2\pi)$, intrinsic quality factor $Q_0$, coupling coefficient $\beta$, and loaded half-bandwidth $\Delta f_{0.5}^L = f_0/(2Q_L)$ giving
\begin{equation}
    |m_r| = \frac{\sqrt{\rho_a c^3 \hbar}}{4\pi\sqrt{2}} \frac{g_{a\gamma\gamma}}{(\beta + 1)}\cdot\frac{1}{\sqrt{1 + (f_a/\Delta f_{0.5}^L)^2}}\cdot\left(\frac{Q_0|\mathscr{H}|}{f_0}\right), \\
    \label{eq:mr}
\end{equation}
with $f_a$ the axion frequency. Equation~\eqref{eq:mr} shows that, for axion frequencies within the loaded resonator bandwidth ($f_a\ll\Delta f_{0.5}^L$), the cavity-dependent contribution to the modulation index is proportional to $Q_0|\mathscr H|/f_0$. For axion frequencies well outside the resonator bandwidth ($f_a\gg\Delta f_{0.5}^L$), the dependence on $Q_0$ and $f_0$ is cancelled by the bandwidth suppression factor, leaving $m_r\propto|\mathscr H|$. Since the experiment is most sensitive to axions whose frequencies lie within the resonator bandwidth, maximising $Q_0|\mathscr H|/f_0$ provides the natural optimisation objective and therefore forms the basis of the composite FoM introduced below.

\subsubsection{Composite Figure of Merit}
\label{sec:FoM}

Each candidate cavity geometry is described by a design vector, $\mathbf{x}$, whose components define the geometric degrees of freedom of the cavity family under consideration. For each candidate geometry, an eigenfrequency study is performed to obtain the corresponding set of cavity eigenmodes, $\{i\}$. The optimisation FoM is therefore defined as
\begin{equation}
    F(\mathbf{x})=
    \frac{\hat{Q}_0^{\mathrm{opt}}\,|\mathscr{H}_{\mathrm{opt}}|}{f_{\mathrm{opt}}},
    \label{eq:FoM}
\end{equation}
where ${\mathrm{opt}}$ denotes the eigenmode satisfying
\begin{equation}
    \mathrm{opt}=
    \arg\max_i
    \left(
        \frac{\hat{Q}_0^i\,|\mathscr{H}_i|}{f_i}
    \right).
\end{equation}
The modulation index may therefore be re-written as
\begin{equation} \label{eq:mamUnit}
    |m_r| = \frac{\sqrt{\rho_a c^3 \hbar}}{4\pi\sqrt{2}} \frac{g_{a\gamma\gamma}}{(\beta + 1)}\cdot\frac{F(\mathbf{x})}{\sqrt{1 + (f_a/\Delta f_{0.5}^L)^2}}.
\end{equation}

This FoM contains all cavity-dependent parameters governing the expected axion-induced modulation index $m_r$ and therefore provides a quantitative metric for optimising and comparing candidate cavity geometries for ULDM detection. We will show in Section~\ref{sec:Measurement} that this FoM is proportional to the signal-to-noise ratio (SNR) of a twisted \remove{anyon}\review{chiral} cavity ULDM experiment, hence higher FoM gives a shorter measurement time to reach the same exclusion limits. In general, the FoM reflects that large volume cavities are more sensitive, as the frequency, $f_i$, will decrease, and the Q-factor will also increase at lower frequencies.

\begin{table*}[!t]
%\small
    \caption{Summary of the performance metrics for the optimal resonant mode, $\mathrm{opt}$, that maximises the optimisation figure of merit, $F(\mathbf{x})$, for each optimised helical resonator considered in this work. }
    \label{tab:FoM_summary}
    \begin{ruledtabular}
        \begin{tabular}{lcccccc}
            \textbf{Resonator} &
            \textbf{$f_{\mathrm{opt}}$ (GHz)} &
            \textbf{$\mathscr{H}_{\mathrm{opt}}$} &
            \textbf{$G_{\mathrm{opt}}$ ($\Omega$)} &
            \textbf{$\hat R_s$ (n$\Omega$)} &
            \textbf{$\hat{Q}_0^{\mathrm{opt}}$ ($\times 10^{12}$)} &
            \textbf{$F(\mathbf{x})$ (s)} \\
            \hline
            SM (optimised) & 3.53 & -0.611 & 736 & 0.630 & 1.17 & 202 \\
            AdM (optimised) & 11.9 & 1.27 & 637 & 170 & 0.00374 & 0.400 \\
            M\"obius benchmark & 17.7 & 0.951 & 342 & 170 & 0.00201 & 0.108 \\
        \end{tabular}
    \end{ruledtabular}
\end{table*}

\subsection{Candidate Cavity Geometries}
\label{sec:cavity_geometries}

The cavity geometries considered in this work are designed to simultaneously maximise the electromagnetic helicity, $\mathscr H$, and intrinsic quality factor, $Q_0$, while remaining compatible with practical superconducting fabrication. Two fabrication approaches are considered to achieve this objective. The first exploits the geometric freedom afforded by additive manufacturing (AdM) to realise continuously mirror-asymmetric cavity boundaries. The second considers cavity geometries compatible with subtractive manufacturing (SM) from high-purity bulk niobium using established superconducting radio-frequency (SRF) cavity fabrication techniques. For both approaches, the admissible design space is constrained to ensure fabrication feasibility and compatibility with operation on the mixing chamber (MXC) plate of a dilution refrigerator.

The M\"obius resonator reported in Ref.~\cite{Bourhill2023}, which represented the highest-performing cavity geometry identified in that study, exemplifies a class of resonators that generate large $\mathscr H$ through continuously mirror-asymmetric cavity boundaries. Such geometries cannot readily be realised using conventional SM and are therefore well suited to AdM. In this work, this design philosophy is further investigated through the optimisation of a twisted ring (TR) resonator, which is based on the edge-free twisted-ring geometry introduced in Ref.~\cite{Paterson2026InverseDesign}. The complete geometric parameterisation and construction procedure for the TR resonator are described in Ref.~\cite{Paterson2026InverseDesign}.

Although AdM provides considerable geometric freedom, the microwave performance of AdM niobium cavities remains limited by fabrication-induced loss mechanisms (see Appendix~\ref{app:Rs_analysis} for a detailed discussion of these mechanisms and their contribution to $R_s$). For the purposes of calculating $\hat Q_0^\mathrm{opt}$ during optimisation, the surface resistance is estimated as
\begin{equation}
    \hat{R}_s = 170~\mathrm{n}\Omega,
\end{equation}
corresponding to the lowest experimentally demonstrated $R_s$ reported for an AdM niobium SRF cavity~\cite{Frigola2015}. Appendix~\ref{app:Rs_3D_Nb_SRF} discusses the applicability and limitations of this value for the AdM resonators considered in this work.

The second fabrication approach considers cavity geometries compatible with conventional SM from high-purity bulk niobium. Although SM imposes greater geometric constraints than AdM, it enables higher $Q_0$ values owing to the exceptionally low microwave surface resistances $R_s$ achievable with bulk niobium. \review{Detailed geometric specifications of the optimised SM resonator are not provided due to the sensitive nature of the design and will be addressed in a separate publication.} To reflect this performance in the optimisation, $\hat{R}_s$ is taken to be
\begin{equation}
    \hat{R}_s = 0.63~\mathrm{n}\Omega,
    \label{eq:Rs_BN}
\end{equation}
corresponding to the lowest experimentally demonstrated $R_s$ reported for a bulk-niobium SRF cavity following medium-temperature vacuum heat treatment~\cite{Posen2020}. This value is adopted as an optimistic estimate for calculating $\hat Q_0^\mathrm{opt}$, although its applicability under the millikelvin-temperature and low-power conditions considered here remains uncertain. Further discussion is provided in Appendix~\ref{app:Rs_hh_Nb_SRF}.

The optimisation is performed independently for each candidate cavity family using the methodology described in Section~\ref{sec:FOMderivation}. A genetic algorithm explores the admissible design space while finite-element eigenmode simulations are used to evaluate $F(\mathbf{x})$ for each candidate geometry. The optimisation parameters are selected independently for each cavity family to ensure adequate search coverage and convergence of the GA.

%%%%%%%%%%%%%%%%%%%%%%%%%%%%%%%%%%%%%%%%%%

\section{Resultant Optimised Cavity Designs}
\label{sec:results}

\subsection{Overview of Optimised Cavity Performance}
\label{sec:results_overview}

Table~\ref{tab:FoM_summary} summarises the figure of merit, $F(\mathbf{x})$, for the optimised resonators considered in this work, together with the key electromagnetic performance metrics of their optimal resonant modes, ${\mathrm{opt}}$. The benchmark M\"obius resonator from Ref.~\cite{Bourhill2023} is included as a heuristic-design reference. The optimised SM resonator achieves a substantially larger $F(\mathbf{x})$ than both the optimised AdM resonator and the benchmark Möbius resonator, primarily owing to the lower $\hat{R}_s$ assumed for subtractive manufacturing.
% ==========================================================

%%%%%%%%%%%%%%%%%%%%%%%%%%%%%%%%%%%%%%%%%%

\section{Discussion}

\label{sec:discussion}

\subsection{Performance Gain Over Heuristic M\"obius-Based Designs}
\label{sec:heuristic_mobius}

\remove{Every optimised cavity geometry considered in this study exhibits a larger $F(\mathbf{x})$ than the heuristic M\"obius resonator reported in Ref.~\cite{Bourhill2023}. The optimised SM resonator achieves the largest improvement, with an $F(\mathbf{x})$ more than three orders of magnitude greater than that of the M\"obius benchmark. Since larger values of $F(\mathbf{x})$ correspond to shorter measurement times for a fixed experimental sensitivity, the optimised resonators are predicted to substantially reduce the measurement time required for a single-mode helical microwave haloscope search for ULDM axions compared with the heuristic M\"obius-based design.}

\review{The optimised AdM cavity achieves a larger $F(\mathbf{x})$ than the heuristic M\"obius resonator reported in Ref.~\cite{Bourhill2023}. The optimised SM resonator achieves the largest $F(\mathbf{x})$ overall, with an $F(\mathbf{x})$ more than three orders of magnitude greater than that of the M\"obius benchmark. This improvement is primarily driven by the substantially lower surface resistance assumed for the bulk-niobium SM approach. Since larger values of $F(\mathbf{x})$ correspond to shorter measurement times for a fixed experimental sensitivity, the resulting SM resonator is predicted to substantially reduce the measurement time required for a single-mode chiral microwave haloscope search for ULDM axions compared with the heuristic M\"obius-based design.}

\subsection{Predicted Performance Advantage of Subtractive Over Additive Manufacturing}
\label{sec:disc_add_vs_sub}

The optimised AdM cavities support resonant modes with larger $\mathscr H$ than the SM cavities considered in this work. This improvement reflects the additional geometric freedom provided by AdM, which enables continuously twisted cavity walls and allows mirror asymmetry to be distributed throughout the entire volume of the resonant mode. 

Despite this advantage, the optimal resonant modes of the optimised SM cavity resonators consistently achieve larger values of $F(\mathbf{x})$. The improvement in $\mathscr H_{\mathrm{opt}}$ provided by the continuously twisted cavity walls is outweighed by the substantially higher $\hat Q_0^{\mathrm{opt}}$ achievable in bulk-niobium SRF cavities. This arises because, under the assumptions adopted in this work, the surface resistance of bulk-niobium cavities is much lower than that of AdM niobium cavities. 

It should be emphasised that this conclusion is specific to the values of $\hat R_s$ and the cavity designs considered in this work. The values adopted for each fabrication method are based on the lowest reported microwave surface resistances and may differ from those ultimately achieved experimentally. Furthermore, the present study compares cavity designs that are naturally suited to AdM and SM, rather than identical cavity geometries fabricated using different manufacturing methods. The results should therefore be interpreted as indicating that, for the assumptions and cavity families considered here, the SM designs provide the highest predicted performance, rather than as a general conclusion regarding the relative merits of AdM and SM. 

\subsection{Planned Experimental Realisation}
\label{sec:disc_fabrication}

The optimised resonators identified in this work will be fabricated using the best available manufacturing approaches representative of those considered throughout the optimisation. The optimised twisted ring (TR) resonator along with the benchmark M\"obius resonator will be fabricated from Nb--47Ti (wt.\%) using laser powder bed fusion (LPBF) with \textit{in situ} alloying by the School of Metallurgy and Materials at the University of Birmingham~\cite{Ersoz2026}. The optimised subtractively manufactured bulk-niobium SRF resonators will be fabricated by the Thomas Jefferson National Accelerator Facility using established SRF cavity manufacturing techniques~\cite{Ciovati2018,Ciovati2007}.

%%%%%%%%%%%%%%%%%%%%%%%%%%%%%%%%%%%%%%%%%%

\section{Detecting Ultralight Dark Matter}
\subsection{Measurement Scheme and SNR for the Detection of Amplitude Modulation Sidebands} \label{sec:Measurement}
\begin{figure*}[!t]
    \centering
    \includegraphics[width=0.75\textwidth]{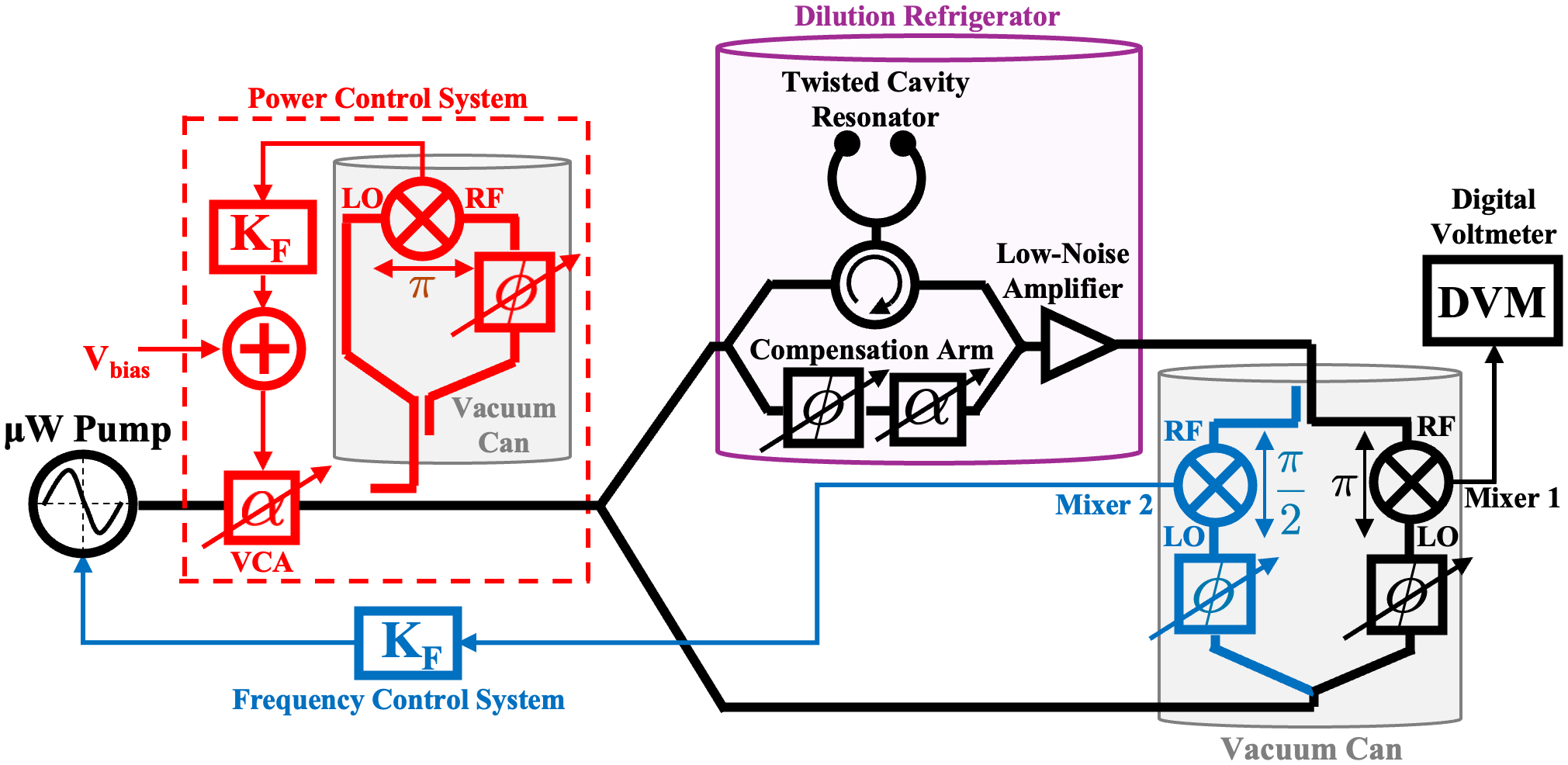}
    \caption{\label{fig:readout} Schematic of the readout system for measurement of amplitude modulation (AM) sidebands induced in the twisted \remove{anyon}\review{chiral} cavity resonator from interaction with a dark matter halo of ultralight axions. Elements in black denote the core interferometric noise measurement system~\cite{ivanov_microwave_1998,rubiola_advanced_2002}, which converts AM sidebands from the resonator into a voltage output at mixer 1. The frequency control system is shown in blue, while the power control system (PCS) is shown in red. In this schematic $\phi$ and $\alpha$ each denote a variable phase shifter or attenuator respectively, $\mathrm{K}_{\mathrm{F}}$ denotes a filtering and gain stage in a feedback control system, and VCA is a voltage-controlled attenuator.}
\end{figure*}

To measure AM-sidebands produced in the twisted \remove{anyon}\review{chiral} cavity, we propose a microwave interferometric noise measurement system~\cite{ivanov_microwave_1998,rubiola_advanced_2002}, which is tuned to measure AM-noise of the device under test (DUT), shown in black in Fig.~\ref{fig:readout}. Our DUT is the single-port twisted \remove{anyon}\review{chiral} cavity resonator of any design discussed in Section~\ref{sec:results}, with a circulator to extract the reflected signal. A second mixer tuned to be phase sensitive (blue in Fig.~\ref{fig:readout}) is used to tightly lock the pump oscillator to the cavity frequency. 

\review{Changes in the frequency difference between the cavity and pump can cause reduction in circulating cavity power, degrade interferometer carrier suppression, and convert frequency to amplitude noise. Additionally, a practical setup will have imperfect tuning to the AM-noise quadrature and retain some phase noise sensitivity. With a narrow-linewidth cavity, a phase-locked loop (PLL) is therefore essential. PLL systems have been designed for operation over several weeks~\cite{lipphardt_optical_2017}, with later systems including PLLs used for 10 months with 19\% downtime~\cite{lipphardt_continuous_2023}. Pound-stabilized cryogenic sapphire oscillators have been reported to operate for nine months~\cite{nand_ultra-stable_2011}. The Thomas Jefferson National Accelerator Facility routinely operates locking to narrow linewidth superconducting niobium cavities~\cite{powers_theory_2005}. Uninterrupted locking is not essential and long-term measurements of weak periodic signals can accommodate interruptions~\cite{stanwix_improved_2006}. With this in mind, we assume a tight lock with allowable interruptions excluded from the effective integration time. We expect this will reduce the effect of cavity-pump detuning below the level of the dominant noise sources considered here.}

\review{In addition, temperature fluctuations can cause differential phase and amplitude mismatch between the interferometer arms. Hence, the interferometer must be made compact, placing the cavity and compensation arm close together in the temperature controlled environment of the dilution refrigerator. This noise contribution will ultimately be determined experimentally by measurement of the noise floor of a compact cryogenic interferometer and temperature-to-amplitude/phase transfer functions.}

\review{For the projected sensitivity here, we consider the dominant noise sources to be pump AM-noise, and system electronics noise including amplifier and mixer. However, in general, vibration of the cavity walls may provide an additional noise source. Vibrations of the cavity walls can be severe for precision experiments involving superconducting cavities~\cite{berlin_axion_2020, cervantes_deepest_2024}. Typically cryogenic vibration will contain a broadband background along with harmonics of the pulse tube, which could in produce AM and PM sidebands, degrading sensitivity or producing spectral lines which appear like a candidate. We don't include an estimate of these effects here as they are highly dependent on the physical cavity, mounting and cryostat. We will experimentally measure this in future work.} 

\remove{We}\review{Here we} define signal-to-noise ratio (SNR) for axion-detection in this system as
\begin{equation} \label{eq:SNRdef}
    SNR = \frac{\delta u_{mix}^r}{\delta u_{n/f}} = \frac{S_{AD}^r \cdot \delta m_r}{\sqrt{(S_{AD}^p \cdot \delta m_p)^2 + (\delta u_{mix}^e)^2}}
\end{equation}
where $\delta u_{mix}^r$ is the rms mixer voltage spectral density due to the axion signal from the resonator, and $\delta u_{n/f}$ is the total noise spectral density competing against the axion signal. We assume the dominant noise sources are \review{uncorrelated} contributions from the AM-noise of the pump oscillator $\delta u_{mix}^p$ and the electronics noise floor of the readout system (mixer and amplifier) $\delta u_{mix}^e$. The interferometric readout system has some sensitivity, $S_{AD}^r$ to amplitude noise from the resonator where $\delta u_{mix}^r = S_{AD}^r \cdot \delta m_r$, with $\delta m_r$ the rms AM-noise spectral density of amplitude modulation produced by the axion signal. Similarly we define the sensitivity to pump AM-noise $S_{AD}^p$ where $\delta u_{mix}^p = S_{AD}^p \cdot \delta m_p$ and $\delta m_p$ is the rms AM-noise spectral density of the microwave pump source. 

To reduce the contribution from pump AM-noise, a power control system (PCS) is employed (Fig.~\ref{fig:readout} in red), where a mixer-based system detects AM-noise in a feedback loop with a voltage-controlled attenuator (VCA) to actively suppress noise in real time. In this way, AM-noise of the pump is reduced to the noise floor of the PCS.

As derived in~\cite{Bourhill2023}, the reflected power in each modulation sideband at modulation frequencies $\omega_0 \pm \omega_a$, can be expressed, in terms of the modulation index $m_r$ (given in ~\eqref{eq:mamUnit}) as
\begin{equation} \label{eq:PrefAM}
    P_{ref}(\omega_0 \pm \omega_a) = P_{inc}\cdot\frac{4 \beta^2}{(\beta+1)^{2}}\cdot (\frac{m_r}{2})^2
\end{equation}
where $P_{inc}$ is the incident power on the resonator and $\beta$ is the coupling coefficient.

The mixer output voltage can be expressed as $u_{mix} = \chi \sqrt{P_{RF}}\cos(\delta\phi_{21})$ where $\chi$ is the mixer conversion efficiency, $P_{RF}$ is the power at the RF port, and $\delta\phi_{21}$ is the phase shift between the RF and LO ports. Hence, when tuning $\delta \phi_{21}$ to the maximally amplitude sensitive region ($\delta\phi_{21} = n\pi,~n\in \mathbb{Z}$), and employing~\eqref{eq:PrefAM} the sensitivity to AM-noise from a dark matter signal in the resonator is
\begin{equation} \label{eq:Sres}
    S_{AM}^r = \frac{\delta u_{mix}^r}{\delta m_r} = 2\cdot\chi \sqrt{\frac{P_{inc}\cdot K_{LNA}}{2}}\cdot\frac{\beta}{\beta+1}
\end{equation}
where $K_{LNA}$ is the gain of the low-noise amplifier (LNA). The division by 2 (inside the square root) accounts for half the power lost to the bright port of the interferometer, and the multiplication by 2 (outside the square root) accounts for the coherent addition of both positive and negative frequency sidebands. Using a similar method (see Appendix~\ref{app:SpumpAM}) we see that the sensitivity to AM-noise of the pump is
\begin{equation} \label{eq:Spump}
  \begin{aligned}
    S_{AM}^p &= \frac{\delta u_{mix}^p}{\delta m_p}\\
    &= \chi\sqrt{\frac{P_{inc} \cdot K_{LNA}}{2}}\cdot\frac{2\beta}{\beta+1}\cdot\frac{|f_a|/\Delta f_{0.5}^L}{\sqrt{1 + (f_a/\Delta f_{0.5}^L)^2}}
  \end{aligned}
\end{equation}
for which we assume that the cancellation arm is tuned to have the same loss and phase shift as the resonator on resonance.

To derive the final SNR we must recast the rms AM-noise spectral density $\delta m_r$ in terms of the modulation index $m_r$ given in~\eqref{eq:mamUnit}. For averaging times $\tau$ less than the axion coherence time we have
$\delta m_r = m_r \cdot \sqrt{\tau/2}$, with standard $\sqrt{\tau}$ dependence and a factor of $1/\sqrt{2}$ to convert to rms fluctuations. For averaging times longer than the axion coherence time we use the standard result to replace $\sqrt{\tau}$ with $(\tau/\Delta f_a)^{1/4}$, where $\Delta f_a$ is the axion linewidth and is equal to $f_a/Q_a$ with $Q_a$ the expected axion quality factor of $10^6$~\cite{sikivie_invisible_2021}. Since our detection is broadband with a large range of axion coherence times, to avoid discontinuities we use the function
\begin{equation} \label{eq:tauDependence}
    \delta m_r = \frac{m_r}{\sqrt{2}}\cdot\sqrt{\frac{\tau}{\sqrt{1 +\tau\Delta f_a}}}
\end{equation}
which interpolates between the two functions for short and long coherence times. Inserting~\eqref{eq:mamUnit},~\eqref{eq:Sres}, and~\eqref{eq:tauDependence} into~\eqref{eq:SNRdef} gives
\begin{equation} \label{eq:SNRfinal}
  \begin{aligned}
    SNR = &\frac{g_{a\gamma\gamma}\sqrt{\rho_a c^3 \hbar}}{4 \pi} \cdot\chi \sqrt{\frac{P_{inc} \cdot K_{LNA}}{2}} \cdot \frac{\beta}{(\beta+1)^{2}}\\
    &\cdot\frac{F(\mathbf{x})}{\sqrt{1 + (f_a/\Delta f_{0.5}^L)^2}}\cdot\sqrt{\frac{\tau}{\sqrt{1+\tau \Delta f_a}}}\cdot\frac{1}{\delta u_{n/f}}
  \end{aligned}
\end{equation}
with the total noise floor given by
\begin{equation} \label{eq:NFfinal}
  \begin{aligned}
    \delta u_{n/f} = \sqrt{(S_{AM}^p\cdot \delta m_p)^2 + (\delta u_{mix}^e)^2}
  \end{aligned}
\end{equation}
with $S_{AM}^p$ given by~\eqref{eq:Spump}. We note that our signal is low-pass filtered, and sensitivity to pump AM-noise ($S_{AD}^p$) is high-pass filtered. For both, the corner frequency is the cavity loaded half-bandwidth. For frequencies $f_a << \Delta f_{0.5}^L$ the pump AM-noise is attenuated, the electronics noise is the dominant noise source, and the $SNR$ is proportional to the cavity FoM ($SNR \propto F(\mathbf{x})$). For frequencies $f_a >>\Delta f_{0.5}^L$, the pump noise dominates, then using~\eqref{eq:FoM} $F(\mathbf{x}) = ((\beta +1)\cdot\mathscr{H})/(2\Delta f_{0.5}^L)$, so we see that $\Delta f_{0.5}^L$ cancels giving
\begin{equation} \label{eq:SNRpump}
    SNR = \frac{g_{a\gamma\gamma}\sqrt{\rho_a c^3 \hbar}}{16 \pi} \cdot\frac{|\mathscr{H}|}{f_a}\cdot\sqrt{\frac{\tau}{\sqrt{1+\tau \Delta f_a}}}\cdot\frac{1}{\delta m_p}
\end{equation}
which is independent of $f_0$ and $Q_0$. 

\subsection{Projected Sensitivity to Ultralight Dark Matter}
\label{sec:ProjSensitivity}
\begin{figure*}[!t]
  \centering
  \subfloat[]{%
    \includegraphics[width=0.49\textwidth]{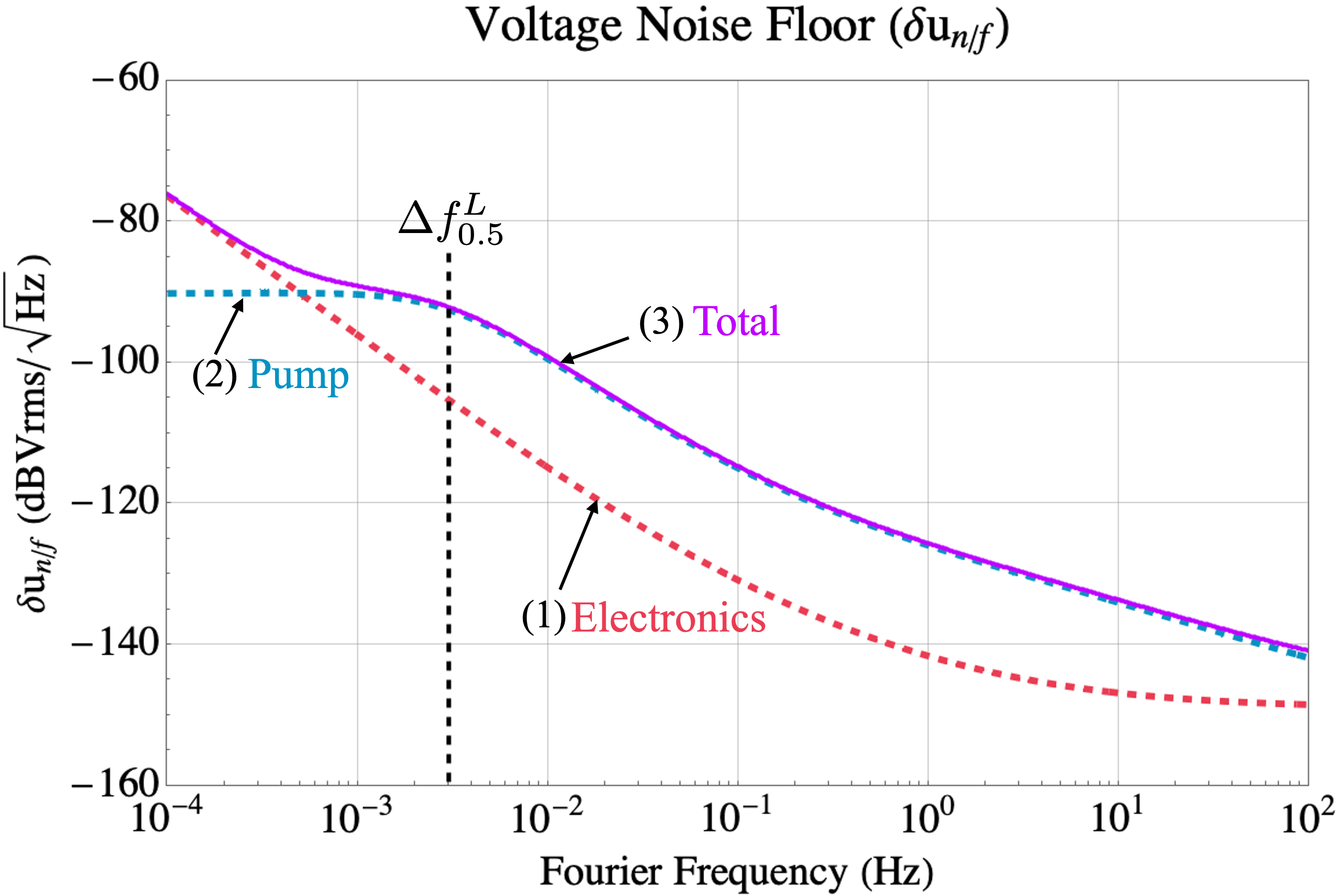}
    \label{fig:vNF}
  }
  \hfill
  \subfloat[]{%
    \includegraphics[width=0.49\textwidth]{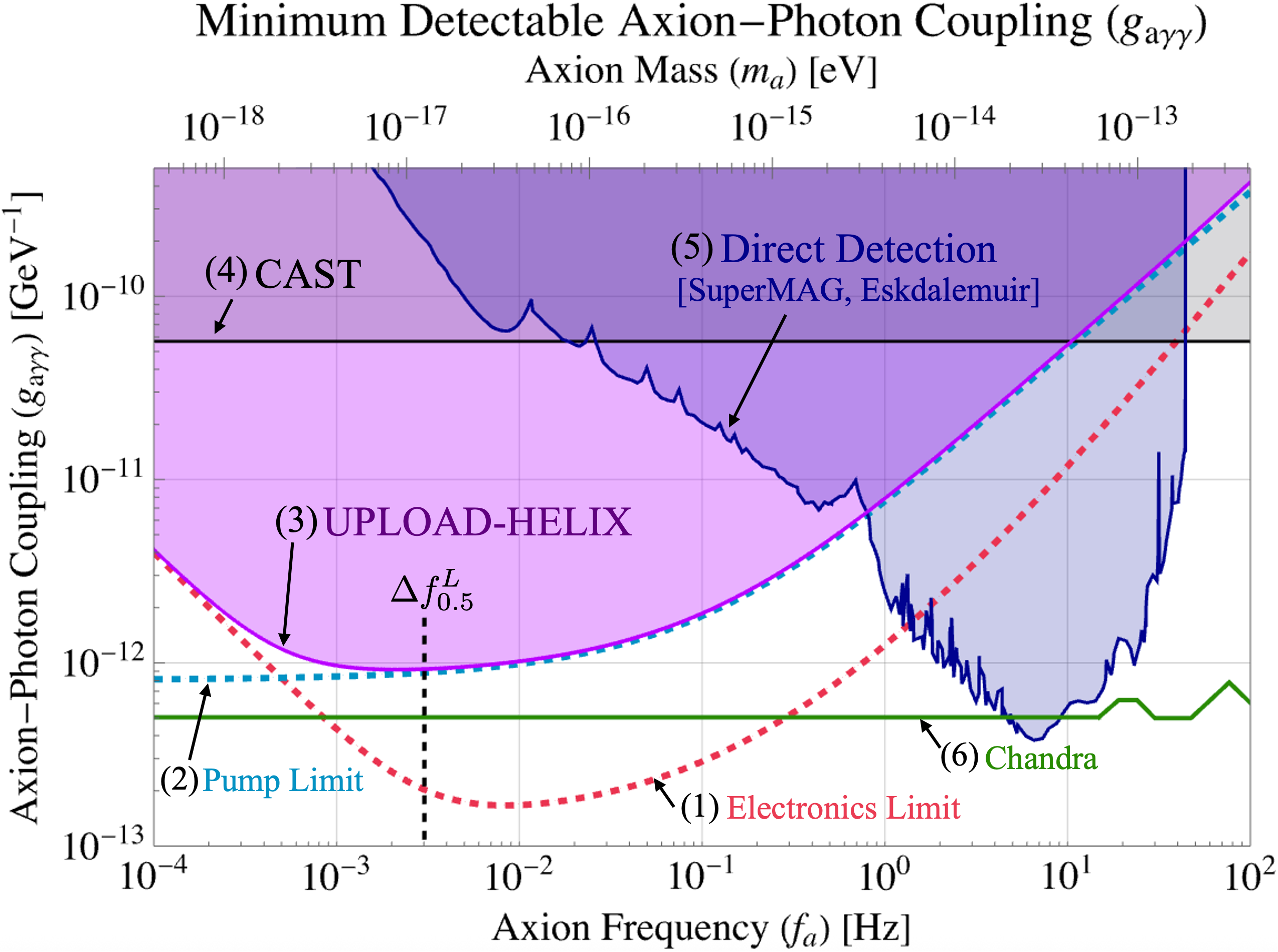}
    \label{fig:gNF}
  }
  \caption{Expected sensitivity for the twisted \remove{anyon}\review{chiral} cavity to ultralight dark matter axions employing~\eqref{eq:SNRfinal} and~\eqref{eq:NFfinal}, setting $SNR=1$, using the optimised SM cavity (see table~\ref{tab:FoM_summary}), assuming readout system parameters in table~\ref{tab:readoutParams}, and using a fit to measured data for electronics noise (see Appendix~\ref{app:mixer}). \textbf{(a)}~Measurement system voltage noise floor. Trace 1 shows the electronics noise floor. Trace 2 shows the expected contribution from pump AM-noise reduced to the PCS noise floor. Trace 3 is the total expected noise floor of the measurement system. \textbf{(b)}~Expected limits on axion-photon coupling that can be set by the experiment. Traces 1 and 2 show limits due solely to electronics or pump noise respectively. Trace 3 shows the projected exclusion region of axion-photon coupling strength that can be searched with the twisted \remove{anyon}\review{chiral} cavity. Trace 4 shows the CAST limit~\cite{altenmuller_new_2024}. \review{Trace 5 shows the combined direct detection limit from SuperMAG~\cite{arza_earth_2022,friel_search_2024}, and Eskdalemuir~\cite{nishizawa_axion_2026}. Trace 6 shows constraints from the Chandra observation of the H1821+643 quasar~\cite{sisk-reynes_new_2021}.}}
  \label{fig:sens}
\end{figure*}

In Fig.~\ref{fig:sens}, we estimated the sensitivity of the twisted \remove{anyon}\review{chiral} cavity to detect ULDM using the optimal FoM of $202~\mathrm{s}$ for the best SM cavity (see table~\ref{tab:FoM_summary}), and assumed standard values for an interferometric noise measurement system shown in table~\ref{tab:readoutParams}. This assumes using readily available technology such as a 32 dB gain LNA from Low-Noise-Factory. Note that the probe coupling $\beta$ can be readily adjusted, and we set $\beta = 1$ as this is the optimal value from~\eqref{eq:SNRfinal}. We then used the voltage noise floor in Fig.~\ref{fig:vNF}, and employing~\eqref{eq:SNRfinal} set $SNR = 1$ and solved for $g_{a\gamma\gamma}$ to arrive at Fig.~\ref{fig:gNF}.

The electronics noise floor (trace 1 in Fig.~\ref{fig:vNF}) is inferred from measurements with a ML10220 mixer-based detector under a vacuum of around 5 mbar from 0.2 mHz to 100 Hz described in appendix~\ref{app:mixer}. The pump AM-noise was calculated from the mixer noise floor divided by the expected PCS sensitivity. This becomes trace 2 in Fig.~\ref{fig:vNF} by multiplication by the sensitivity of the interferometric readout system to pump AM-noise in~\eqref{eq:Spump}. Using~\eqref{eq:NFfinal} the total noise is shown in Fig.~\ref{fig:vNF}, trace 3. Due to the high pass filter on pump-noise sensitivity we are limited by the pump outside the resonator bandwidth, and the electronics inside the bandwidth. 

In Fig.~\ref{fig:gNF}, traces 1, 2, and 3 are conversions from each corresponding voltage noise floor to minimum detectable $g_{a\gamma\gamma}$, with the final expected sensitivity shown in trace 3. At higher frequencies we lose sensitivity outside the resonator bandwidth, and at the lower frequencies we are limited by increasing mixer flicker noise. \remove{Trace 4 shows the best direct detection limits in the mass range set by CAST.} \review{Traces 4, 5, and 6 show current limits in the mass range from helioscopes~\cite{altenmuller_new_2024}, direct detection~\cite{arza_earth_2022,friel_search_2024,nishizawa_axion_2026}, and astrophysical observations from the Chandra mission~\cite{sisk-reynes_new_2021}.} \remove{With a conservative readout system}\review{Under the above inferred electronics noise, the assumptions in Sec.~\ref{sec:Measurement}, the optimised SM cavity in table~\ref{tab:FoM_summary}, system parameters in table~\ref{tab:readoutParams}} and 3 months averaging time, we expect to set the most sensitive direct detection limits in the mass range, reaching below $10^{-11}~\mathrm{GeV}^{-1}$ over more than 4 orders of magnitude in mass from $4\times10^{-19}~\mathrm{eV}$ to $4\times10^{-15}~\mathrm{eV}$.

\begin{table}[!t] 
%\small % Change table font size
    \caption{Summary of readout system parameters used to produce plots in figure~\ref{fig:gNF}.}
    \label{tab:readoutParams}
    \begin{ruledtabular}
        \begin{tabular}{cc}
            \textbf{Parameter} &
            \textbf{Assumed Value} \\
            \hline
            $\chi~[\mathrm{V}\sqrt{\mathrm{W}}]$ & 7.5 \\
            $P_{inc}~[\mathrm{dBm}]$ & -15\\
            $K_{LNA}~[\mathrm{dB}]$& 32 \\
            $\beta$ & 1 \\
            $\tau~[\mathrm{months}]$ & 3\\
            PCS Sensitivity $(\mathrm{V})$ & 1.9 \\
        \end{tabular}
    \end{ruledtabular}
\end{table}

The electronic noise limit in Fig.~\ref{fig:gNF} may be improved by either increasing incident power on the resonator ($P_{inc}$), or amplifier gain ($K_{amp}$). \remove{The maximum incident power will be limited by the cooling power of the dilution refrigerator.}\review{At critical coupling ($\beta=1$), the assumed $P_{\mathrm{inc}}=-15~\mathrm{dBm}$ corresponds to $P_{\mathrm{diss}}=31.6~\mu\mathrm{W}$ within the resonator. The maximum microwave power will therefore be limited by the cooling power of the dilution refrigerator required to maintain the target operating temperature of $T=15~\mathrm{mK}$. The compatibility of this operating point with the assumed surface resistance of $\hat R_s=0.63~\mathrm{n}\Omega$~\eqref{eq:Rs_BN} is discussed in Appendix~\ref{app:Rs_hh_Nb_SRF}.} Amplifier gain could be increased with either a choice of a higher gain amplifier, or potentially two amplifiers in series, provided the second amplifier can be kept out of saturation with correct interferometer balancing. This can be done because below 10 Hz we expect electronics noise to be dominated by the mixer not amplifier. As total amplifier gain increases we'd expect this corner frequency to decrease, as more of the spectrum eventually becomes dominated by amplifier noise. However, from~\eqref{eq:SNRpump} neither will have any effect on the limit due to pump noise. Novel possibilities will need to be explored to surpass this limit. Some are briefly introduced below, but details are left to future works.
\begin{enumerate}
    \item One possible method is to reduce the pump noise further by introducing an interferometric component to the PCS. This would require a second interferometer, which is exclusively part of the PCS. An interferometer increases sensitivity by the gain of the low-noise amplifier, allowing orders of magnitude reduction in the minimum detectable noise. This is a well-known technique for reducing phase noise~\cite{ivanov_microwave_1998}, though little attention has been given to application to amplitude noise. 
    \item Another method would be to cancel the pump noise in the interferometer. Due to the filtering effect of the resonator the interferometer only cancels pump noise at the resonant frequency, but if AM-sidebands of the pump experienced the same magnitude and phase response in both arms they would be cancelled. One possibility is to employ two modes of opposite helicity in the cavity, each excited by a different arm of the interferometer. AM-sidebands induced by the axion will experience a $\pi$ phase shift relative to the carrier when the sign of the helicity is changed~\cite{goryachev_axion_2019}. Hence, when summing these two arms at the interferometer output, the carrier and its modulation sidebands could be cancelled while the axion modulation sidebands are added. 
\end{enumerate}

\section{Conclusion}

An inverse-design framework was developed to optimise superconducting niobium \remove{helical}\review{chiral} cavity resonators for single-mode ultralight dark matter axion searches using a figure of merit derived to minimise the measurement time required to reach a fixed experimental sensitivity. The optimisation was applied to candidate cavity geometries representative of both additive and subtractive manufacturing approaches while maintaining compatibility with established superconducting radio-frequency cavity fabrication techniques. \remove{The results indicate that the subtractively manufactured resonator provides the highest predicted performance owing to its substantially higher achievable intrinsic quality factor, yielding the highest figure of merit, $F(\mathbf{x})=202~\mathrm{s}$, representing a substantial improvement over previous helical cavity haloscope designs.}\review{The results identify subtractive manufacturing as the more favourable fabrication approach, primarily due to the substantially lower surface resistance achievable with bulk niobium. Optimisation of the subtractively manufactured resonator geometry yields the highest figure of merit, $F(\mathbf{x})=202~\mathrm{s}$, representing a substantial improvement over the M\"obius resonator proposed for the previous single-mode chiral cavity haloscope.}

A microwave interferometric readout architecture was proposed to transduce the axion-induced amplitude-modulation sidebands generated within the optimised cavity into a measurable voltage signal. When implemented with the optimised twisted \remove{anyon}\review{chiral} cavity, the proposed haloscope is projected to probe axion-photon couplings below $10^{-11}~\mathrm{GeV}^{-1}$ over four orders of magnitude in axion mass, from $4\times10^{-19}$ to $4\times10^{-15}~\mathrm{eV}$, with a peak projected sensitivity down to coupling coefficients of \remove{$2\times10^{-12}~\mathrm{GeV}^{-1}$}\review{$1\times10^{-12}~\mathrm{GeV}^{-1}$} between $4\times10^{-18}$ and $4\times10^{-17}~\mathrm{eV}$. This projected sensitivity extends beyond current exclusion limits from direct detection experiments throughout the accessible mass range.

This work provides a practical pathway towards the experimental realisation of superconducting single-mode \remove{helical}\review{chiral} haloscopes for ultralight dark matter axion searches. Future work will focus on the fabrication and experimental characterisation of the optimised superconducting cavity designs in a dilution refrigerator, implementation of the proposed microwave interferometric readout architecture, and the first experimental search for ultralight dark matter axions using a superconducting single-mode \remove{helical}\review{chiral} haloscope.

%%%%%%%%%%%%%%%%%%%%%%%%%%%%%%%%%%%%%%%%%%

\vspace{6pt} 

\begin{acknowledgements}
The authors would like to thank Dr. Nabin Raut who helped facilitate collaboration between the Quantum Technologies and Dark Matter Research Laboratory and the Thomas Jefferson National Accelerator Facility.

This research was supported by the Australian Research Council Centre of Excellence in Dark Matter Particle Physics under Grant CE200100008. Maxim Goryachev is supported by the Australian Research Council Future Fellowship FT250100168. The work at Jefferson Lab is supported by the U.S. Department of Energy, Office of Science, Office of Nuclear Physics under Contract No. 89243126CSC000213.
\end{acknowledgements}

%%%%%%%%%%%%%%%%%%%%%%%%%%%%%%%%%%%%%%%%%%
\appendix

\section[\appendixname~\thesection]{Surface-Resistance Analysis of Niobium Superconducting Cavities at Millikelvin Temperatures}
\label{app:Rs_analysis}

The aim of this appendix is to determine the expected frequency dependence of $R_s$ for superconducting niobium cavities under the operating conditions considered in this work and to establish the values of $\hat R_s$ used to estimate the intrinsic quality factors, $\hat Q_0^i$, of the cavity eigenmodes required by the optimisation algorithm to evaluate $F(\mathbf{x})$. The analysis first reviews the potentially dominant microwave-loss mechanisms and their associated frequency dependence before establishing the values of $\hat R_s$ adopted for SM high-purity bulk-niobium SRF cavities and AdM niobium SRF cavities.

\subsection[\appendixname~\thesubsection]{Operating Regime}
\label{app:Rs_operating_regime}

Throughout this work, only resonant modes with frequencies in the range
\begin{equation}
    1~\mathrm{GHz}\lesssim f\lesssim20~\mathrm{GHz},
    \label{eqn:freq_range}
\end{equation}
are considered. The cavities are assumed to operate at temperatures of approximately
\begin{equation}
    T\approx15~\mathrm{mK},
    \label{eq:temp_range}
\end{equation}
and under low incident microwave powers,
\begin{equation}
    -30~\mathrm{dBm}\lesssim P_{\mathrm{inc}}\lesssim-10~\mathrm{dBm}.
    \label{eq:power_range}
\end{equation}
\remove{These operating conditions correspond to the operating regime of the ULDM axion search investigated in this work, where millikelvin temperatures suppress thermal noise and measurements are performed at low microwave powers to avoid perturbing the weak axion-induced signal.}\review{These conditions correspond to the operating regime considered for the proposed ULDM axion search, with the maximum incident microwave power ultimately constrained by the available cooling power of the dilution refrigerator and any power-dependent degradation of $R_s$.}

\subsection[\appendixname~\thesubsection]{Surface-Resistance Model}
\label{app:Rs_model}

The $R_s$ of a superconducting cavity may be written as~\cite{Gurevich2012}
\begin{equation}
    R_s(f,T)=R_{\mathrm{BCS}}+R_i,
    \label{eq:Rs_decomp}
\end{equation}
where $R_{\mathrm{BCS}}$ is the thermally activated $R_s$ arising from quasiparticle dissipation described by Bardeen-Cooper-Schrieffer (BCS) theory, and $R_i$ is the residual surface resistance. $R_i$ encompasses all remaining microwave-loss mechanisms that persist after $R_{\mathrm{BCS}}$ has been suppressed. Under the operating conditions considered in this work, these mechanisms are expected to include trapped magnetic flux, two-level systems (TLS), sub-gap states, seam and contact losses, surface roughness, and other fabrication-dependent imperfections.

Compared with conventionally manufactured high-purity bulk-niobium SRF cavities, AdM niobium SRF cavities may exhibit larger residual losses owing to reduced RRR, increased porosity, lower density, greater surface roughness, and microstructural imperfections. These characteristics can enhance several residual-loss mechanisms, including flux-pinning losses, current-crowding effects, and defect-related dissipation, and are consistent with the areas identified for further improvement by Frigola \textit{et al.}~\cite{Frigola2015}.

The relative magnitude and expected frequency dependence of each residual-loss mechanism are examined in the following sections.

\subsection[\appendixname~\thesubsection]{BCS Surface Resistance}
\label{app:Rs_BCS}

For niobium SRF cavities operated at temperatures well below their respective superconducting transition temperatures,
\begin{equation}
    T\ll T_c,
\end{equation}
where $T_c$ is approximately $9.2~\mathrm{K}$ for bulk high-purity ($RRR\approx 300$) niobium~\cite{Gurevich2012} and approximately $9.0~\mathrm{K}$ for electron-beam-melted (EBM)-fabricated niobium~\cite{Frigola2015}. For microwave frequencies satisfying
\begin{equation}
    \hbar\omega\ll\Delta,
\end{equation}
where $\hbar$ is the reduced Planck constant, $\omega=2\pi f$ is the angular frequency and $\Delta$ is the superconducting energy gap, together with peak RF magnetic fields well below the thermodynamic critical field, $B_c$,
\begin{equation}
    B_p\ll B_c,
\end{equation}
the microwave $R_s$ is described by the low-field Mattis-Bardeen approximation~\cite{Gurevich2012},
\begin{equation}
    R_{\mathrm{BCS}}
    =
    \frac{\mu_0^2\omega^2\lambda^3\sigma_n\Delta}
         {k_BT}
    \ln\!\left(
        \frac{C_1k_BT}
             {\hbar\omega}
    \right)
    \exp\!\left(
        -\frac{\Delta}{k_BT}
    \right),
    \label{eq:MB}
\end{equation}
where $\mu_0$ is the permeability of free space, $\lambda$ is the London penetration depth, $\sigma_n$ is the normal-state electrical conductivity, $k_B$ is Boltzmann's constant, $C_1=4/e^\gamma\approx2.246$, and $\gamma\approx0.5772$ is the Euler-Mascheroni constant. The operating conditions defined in Section~\ref{app:Rs_operating_regime} for the ULDM axion search considered in this work satisfy the assumptions required for Eq.~(\ref{eq:MB}) to apply. In practice, this expression is commonly approximated by the semi-empirical form~\cite{Gurevich2012}
\begin{equation}
    R_{\mathrm{BCS}}
    =
    \left(
        \frac{Af^2}{T}
    \right)
    \exp\!\left(
        -\frac{\beta T_c}{T}
    \right),
\end{equation}
where $A$ is a material-dependent constant and $\beta\approx1.9$ for niobium.

For the operating temperatures considered here
\begin{equation}
    \exp\!\left(
        -\frac{\beta T_c}{T}
    \right)
    =
    \exp(-1165)
    \approx0.
\end{equation}
The $R_{\mathrm{BCS}}$ is therefore exponentially suppressed, such that
\begin{equation}
    \boxed{R_{\mathrm{BCS}}\approx0.}
\end{equation}
Consequently, throughout the remainder of this appendix the microwave $R_s$ is assumed to be dominated by residual-loss mechanisms,
\begin{equation}
    \boxed{R_s \approx R_i, \qquad T \approx 15~\mathrm{mK}.}
\end{equation}

\subsection[\appendixname~\thesubsection]{Residual Resistance Mechanisms}
\label{app:Rs_residual}

Having established that $R_{\mathrm{BCS}}$ is negligible under the operating conditions considered in this work, the microwave $R_s$ is expected to be dominated by residual-loss mechanisms. The dominant contributions are expected to be
\begin{equation}
    R_i
    =
    R_{\mathrm{flux}}
    +
    R_{\mathrm{TLS}}
    +
    R_{\mathrm{subgap}}
    +
    R_{\mathrm{seam}}
    +
    R_{\mathrm{rough}}
    + \cdots,
    \label{eq:residual_decomp}
\end{equation}
where $R_{\mathrm{flux}}$ represents losses arising from trapped magnetic flux, $R_{\mathrm{TLS}}$ represents losses due to two-level systems (TLS), $R_{\mathrm{subgap}}$ represents losses associated with sub-gap states, $R_{\mathrm{seam}}$ represents seam and contact losses, and $R_{\mathrm{rough}}$ represents losses arising from surface roughness and current crowding. The contribution of each mechanism to the residual surface resistance, together with its expected frequency dependence, is discussed below.

\subsubsection[\appendixname~\thesubsubsection]{Trapped-Flux Losses}
\label{app:Rs_flux}

Trapped magnetic flux contributes an additional $R_i$ given by~\cite{Dhakal2020}
\begin{equation}
    \begin{aligned}
        R_{\mathrm{flux}}
        &=
        S(f)B_{\mathrm{trapped}} \\
        &=
        S(f)\eta_tB_n,
    \end{aligned}
    \label{eq:flux_loss}
\end{equation}
where $S(f)$ is the trapped-flux sensitivity, $B_{\mathrm{trapped}}$ is the magnetic flux trapped within the superconductor during cooldown, $B_n$ is the ambient DC magnetic field present during the superconducting transition, and $\eta_t$ is the flux-trapping efficiency. The trapped-flux sensitivity $S(f)$ describes the increase in $R_i$ per unit of trapped magnetic flux. The sensitivity depends on the dynamics of trapped vortices driven by RF currents and is influenced by the operating frequency, vortex pinning characteristics, and material properties such as the electron mean free path~\cite{Dhakal2020}. Consequently, trapped magnetic flux can introduce a frequency-dependent contribution to the $R_s$. 

Throughout this work, however, the cavity optimisation assumes operation under idealised experimental conditions, including effective magnetic shielding and active magnetic-field compensation, such that
\begin{equation}
    B_n \approx 0.
\end{equation}
Consequently, the amount of trapped magnetic flux is also expected to be negligible,
\begin{equation}
    B_{\mathrm{trapped}} = \eta_t B_n \approx 0,
\end{equation}
and therefore
\begin{equation}
    R_{\mathrm{flux}}
    \approx0.
\end{equation}
Under these conditions, trapped-flux losses are not expected to contribute significantly to either the magnitude or frequency dependence of $R_s$.

For AdM niobium cavities, $R_{\mathrm{flux}}$ may be more difficult to suppress because fabrication-induced defects, including reduced RRR, increased porosity, reduced density, and microstructural disorder, can provide additional vortex-pinning centres and increase the trapped-flux sensitivity~\cite{Frigola2015}. Consequently, if residual magnetic fields are present during cooldown, $R_{\mathrm{flux}}$ may exceed that of conventionally fabricated high-purity bulk-niobium cavities, leading to a larger frequency-dependent contribution to $R_s$. Nevertheless, effective magnetic shielding, active magnetic-field compensation, and optimised cooldown procedures may still reduce trapped-flux losses to a negligible level.

\subsubsection[\appendixname~\thesubsubsection]{Two-Level-System Losses}
\label{app:Rs_TLS}

A commonly used model for TLS-induced microwave loss is~\cite{Pappas2011}
\begin{equation}
    \frac{1}{Q_{\mathrm{TLS}}}=\frac{F \delta_{\mathrm{TLS}}^0\tanh\left(\frac{\hbar\omega}{2k_B T}\right)}{\sqrt{1+\left(\frac{E}{E_c}\right)^2}},
    \label{eq:QTLS}
\end{equation}
where $F$ is the filling factor of the TLS-hosting medium within the resonator, $\delta_{\mathrm{TLS}}^0$ is the intrinsic TLS loss tangent, $E$ is the microwave electric-field amplitude, and $E_c$ is the characteristic saturation field.

For the millikelvin temperatures considered here, the thermal population factor
\begin{equation}
    \tanh\left(\frac{\hbar\omega}{2k_B T}\right).
\end{equation}
is close to unity throughout most of the $1$-$20~\mathrm{GHz}$ frequency range. Consequently, TLS losses are not expected to exhibit a strong intrinsic frequency dependence, although a small variation may occur at the lowest frequencies where $\hbar\omega$ becomes comparable to $k_BT$.

The low incident microwave powers considered in this work also influence the magnitude of the TLS loss. Equation~\ref{eq:QTLS} predicts that TLS losses become more prominent when the TLS ensemble is not saturated by the applied microwave field. Since the electric-field amplitude scales as
\begin{equation}
        E \propto \sqrt{P_{\mathrm{inc}}},
\end{equation}
where $P_{\mathrm{inc}}$ is the incident microwave power, reducing the input power decreases TLS saturation and therefore increases the associated microwave loss.

Experimental studies have shown that oxide-related TLS losses can contribute significantly to the residual surface resistance, $R_i$, of niobium SRF resonators, although their magnitude depends strongly on the surface preparation and oxide structure~\cite{Romanenko2017,Bafia2024}. Because these losses originate primarily from defects within native oxide layers and interfaces rather than the superconducting niobium itself, surface treatments such as electropolishing, heat treatment, and oxide engineering may substantially reduce their contribution~\cite{Romanenko2017,Bafia2024}. Whether they can be suppressed sufficiently under the low-power operating conditions considered in this work remains uncertain and must ultimately be determined experimentally.

Compared with conventionally fabricated high-purity bulk-niobium cavities, AdM niobium cavities may exhibit larger TLS losses. The increased surface roughness associated with AdM can increase the participation of native oxide layers, while lower RRR, increased porosity, reduced density, and microstructural disorder introduce additional defects and interfaces capable of hosting TLS. Consequently, while the intrinsic frequency dependence of TLS losses is expected to remain weak over the $1$-$20~\mathrm{GHz}$ frequency range considered here, the magnitude of the TLS losses may be larger in AdM niobium cavities. The extent of this enhancement depends on the specific fabrication and post-processing route employed and must ultimately be determined experimentally.

\subsubsection[\appendixname~\thesubsubsection]{Sub-gap-State Losses}
\label{app:Rs_subgap}

Sub-gap states have been proposed as a possible source of residual surface resistance in superconducting niobium by enabling residual quasiparticle dissipation at temperatures where the conventional $R_{\mathrm{BCS}}$ is negligible. They correspond to a finite density of electronic states within the superconducting energy gap, allowing microwave absorption through quasiparticle excitations that are absent in an ideal superconductor~\cite{Gurevich2012}. The frequency dependence and overall contribution of sub-gap states to the $R_i$ are not yet fully understood. Since sub-gap states arise from deviations from the ideal superconducting density of states, their density is expected to depend sensitively on material purity, fabrication history, and surface condition. Consequently, losses associated with sub-gap states may be reduced through optimisation of the cavity fabrication and surface-treatment procedures, although the extent of this reduction has not yet been established conclusively.

Compared with conventionally fabricated high-purity bulk-niobium cavities, AdM niobium cavities typically exhibit lower RRR, increased porosity, reduced density, and greater microstructural disorder. These characteristics may increase the density of sub-gap states and therefore increase the associated microwave loss. However, the magnitude of this contribution remains uncertain and must ultimately be determined experimentally.

In summary, sub-gap-state losses may contribute to the $R_i$ under the operating conditions considered here. Their intrinsic frequency dependence is presently not well understood, although no strong frequency dependence is expected over the $1$-$20~\mathrm{GHz}$ range considered in this work under ideal operating conditions.

\subsubsection[\appendixname~\thesubsubsection]{Seam and Contact Losses}
\label{app:Rs_seam}

Seam and contact losses arise from finite electrical conductance across mechanical joints and interfaces. Their magnitude depends on the RF current crossing the seam and is therefore strongly mode dependent. In the ideal limit of a perfectly conducting seam, the associated contribution to the $R_i$ is negligible~\cite{Brecht2015},
\begin{equation}
    R_{\mathrm{seam}}\approx0.
\end{equation}
Consequently, seam and contact losses are not expected to contribute significantly to either the magnitude or frequency dependence of the $R_s$ under the idealised operating conditions assumed throughout this work. In practical cavities, however, imperfect electrical contacts may introduce an additional residual loss that depends on the overlap between the RF surface-current distribution and the seam location~\cite{Brecht2015}. This contribution is therefore strongly geometry and mode dependent rather than being an intrinsic material property. 

\subsubsection[\appendixname~\thesubsubsection]{Surface Roughness}
\label{app:Rs_roughness}

Surface roughness modifies the local electromagnetic fields and microwave surface-current distribution within the superconducting penetration depth, thereby increasing the effective $R_i$~\cite{Xu2018}. This additional loss is particularly important for AdM cavities, where the as-fabricated surface roughness is generally greater than that of conventionally machined bulk-niobium cavities. Consistent with this, Frigola \textit{et al.} identified reducing surface roughness as an important area for further development of AdM niobium SRF cavities~\cite{Frigola2015}. The magnitude of the resulting loss depends on the local surface topography and may be reduced substantially through mechanical polishing, chemical polishing, or electropolishing. Consequently, roughness-induced losses depend strongly on the fabrication and post-processing route employed.

% Surface roughness modifies the local electromagnetic fields and microwave surface-current distribution within the superconducting penetration depth, thereby increasing the effective $R_i\cite{Xu2018}$. This additional loss is particularly important for AdM cavities, where the as-fabricated surface roughness is generally greater than that of conventionally machined bulk-niobium cavities. Consistent with this, Frigola et al. identified reducing the surface roughness as an important area for further development of AdM niobium SRF cavities\cite{Frigola2015}.

% Surface roughness may therefore be regarded as an additional effective contribution to the $R_i$. The magnitude of this contribution depends on the local surface topography and may be reduced substantially through mechanical polishing, chemical polishing, or electropolishing. The overall roughness-induced loss therefore depends strongly on the fabrication and post-processing route employed.

Although surface roughness may increase the magnitude of the $R_i$, it is not expected to introduce a strong intrinsic frequency dependence. Xu \textit{et al.} showed that roughness-induced losses are governed primarily by the London penetration depth, which is approximately independent of frequency for superconducting niobium~\cite{Xu2018}. Surface roughness is therefore expected to increase the $R_i$ while contributing only weakly to its frequency dependence over the $1$-$20~\mathrm{GHz}$ frequency range considered in this work.

\subsection[\appendixname~\thesubsection]{Expected Frequency Dependence}
\label{app:freq_dep_Rs}

Measurements of high-purity bulk-niobium SRF cavities show that the microwave surface resistance, $R_s$, exhibits a frequency dependence~\cite{Raut2023}. The physical origin of this frequency dependence, however, remains uncertain and is thought to arise from residual-loss mechanisms such as trapped magnetic flux, sub-gap states, and other material imperfections. The preceding analysis indicates that, under the idealised operating conditions adopted throughout this work, including millikelvin temperatures, effective magnetic shielding and active magnetic-field compensation during cooldown, fully welded cavity structures, and optimised surface preparation, these residual-loss mechanisms are either expected to be negligible or to exhibit only a weak intrinsic frequency dependence. In particular, $R_{\mathrm{BCS}}$ is exponentially suppressed at $15~\mathrm{mK}$, while trapped-flux and seam losses are expected to be negligible. Although TLS, surface roughness, and sub-gap states may still contribute to the residual surface resistance, none is presently expected to introduce a strong intrinsic frequency dependence over the $1$–$20~\mathrm{GHz}$ frequency range considered in this work. Accordingly, the experimentally observed frequency dependence of $R_s$ is not expected to be significant under the operating conditions considered here, and the surface resistance is therefore approximated as
\begin{equation}
\boxed{R_s(f)\approx R_i \approx \mathrm{constant}.}
\end{equation}
This approximation forms the basis for treating $R_s$ as constant when estimating $\hat{Q}_0^i$ in the optimisation algorithm.

If residual magnetic fields are present during cooldown, trapped-flux losses may become a significant source of frequency dependence. This effect may be particularly important for AdM niobium cavities, where the increased material disorder associated with current AdM processes is expected to enhance flux trapping.

The following sections establish the values of the microwave surface resistance, $\hat{R}_s$, adopted for conventionally fabricated high-purity bulk-niobium SRF cavities and AdM niobium SRF cavities.

\subsection[\appendixname~\thesubsection]{Surface-Resistance Estimate for High-Purity Bulk Niobium}
\label{app:Rs_hh_Nb_SRF}

For conventionally fabricated high-purity bulk-niobium SRF cavities with RRR $>200$~\cite{Ciovati2018,Ciovati2007}, Posen \textit{et al.}~\cite{Posen2020} reported a $R_s$ of
\begin{equation}
    R_s = 0.63 \pm 0.06~\mathrm{n}\Omega,
\end{equation}
following electropolishing and medium-temperature vacuum heat treatment of a $1.3~\mathrm{GHz}$ \review{single-cell bulk-niobium TESLA} cavity operated below $1.5~\mathrm{K}$ at accelerating gradients up to approximately $16~\mathrm{MV\,m^{-1}}$~\cite{Posen2020}. This is among the lowest experimentally demonstrated \remove{residual surface resistances}\review{$R_i$} reported for a bulk-niobium SRF cavity \remove{and is therefore adopted in this work as an optimistic estimate of highly optimised bulk-niobium SRF performance}.\review{The applicability of this benchmark to the $P_{inc}$ considered for the proposed ULDM axion search is supported by measurements of the same cavity by Romanenko \textit{et al.}~\cite{Romanenko2017}.}

\review{Romanenko \textit{et al.}~\cite{Romanenko20} characterised the cavity in a dilution refrigerator over a broad range of RF-field amplitudes at temperatures down to $10$--$20~\mathrm{mK}$. At $1.3~\mathrm{K}$, the cavity exhibited similarly high $Q_0$, and no measurable RF-field-amplitude dependence of $Q_0$ was observed. This investigated field range corresponds to average cavity photon numbers, $\bar{n}$, extending from approximately $10^{20}$ down to the few-photon regime. To directly compare this experimentally investigated regime with the operating point assumed in our projected sensitivity, we express $P_{\mathrm{inc}}=-15~\mathrm{dBm}$ in terms of $\bar{n}$.}

\review{At resonant critical coupling ($\beta=1$), the incident microwave power is fully dissipated within the resonator, such that $P_{\mathrm{inc}}=-15~\mathrm{dBm}$ corresponds to $P_{\mathrm{diss}}=31.6~\mu\mathrm{W}$. Using $Q_0=\omega U/P_{\mathrm{diss}}$ and $\bar{n}=U/(\hbar\omega)$~\cite{Romanenko2017}, the corresponding average number of photons stored in the cavity is}
\begin{equation}
    \review{\bar{n}=\frac{Q_0P_{\mathrm{diss}}}{\hbar\omega^2}.}
\end{equation}
\review{Using the optimised SM resonator parameters reported in Table~\ref{tab:FoM_summary}, $f_0=3.53~\mathrm{GHz}$ and $Q_0=1.17\times10^{12}$, we obtain $\bar{n}\simeq7.1\times10^{20}$. This is of the same order of magnitude as the upper photon-number regime discussed by Romanenko \textit{et al.}~\cite{Romanenko20}, providing experimental support that the ultralow $R_s$ reported by Posen \textit{et al.}~\cite{Posen2020} is compatible with the $P_{inc}$ of the magnitude assumed in our sensitivity projections.}

\review{However, a surface resistance of $R_s=0.63~\mathrm{n}\Omega$ has not been directly demonstrated at $T\approx15~\mathrm{mK}$ and the $P_{\mathrm{inc}}=-15~\mathrm{dBm}$ operating point simultaneously. Romanenko \textit{et al.} observe a reduction in $Q_0$ below approximately $1~\mathrm{K}$, attributed to TLS-induced dissipation, although this contribution is substantially reduced following vacuum heat treatment~\cite{Romanenko2017}. Achieving an $R_s$ of this magnitude under the proposed operating conditions therefore requires effective suppression of the residual-loss mechanisms discussed in the preceding sections, including trapped magnetic flux, seam losses, surface roughness, TLS-related losses, sub-gap states, and other fabrication-induced imperfections. However, whether TLS and other residual losses can be sufficiently suppressed at approximately $15~\mathrm{mK}$ remains to be established experimentally.}

\review{Maintaining the target operating temperature under the anticipated dissipated microwave power will also depend on the available cooling power of the dilution refrigerator and the thermalisation of the resonator. The measurements of Romanenko \textit{et al.}~\cite{Romanenko20} at temperatures down to $10$--$20~\mathrm{mK}$ provide experimental support for the feasibility of this operating regime. Nevertheless, whether $R_s=0.63~\mathrm{n}\Omega$ can be maintained at approximately $15~\mathrm{mK}$ and at $P_{inc}=-15~\mathrm{dBm}$ considered here ultimately remains to be established experimentally.}

\review{For the purposes of calculating $\hat Q_0^{\mathrm{opt}}$, we therefore adopt the value reported by Posen \textit{et al.}~\cite{Posen2020} as an optimistic, experimentally motivated benchmark for conventionally fabricated high-purity bulk-niobium SRF cavities, rather than as a $R_s$ that has been experimentally demonstrated under the exact operating conditions proposed here. Accordingly, the $\hat{R}_s$ used throughout this work for SM SRF cavities is}
\begin{equation}
    \review{\boxed{\hat{R}_s = 0.63~\mathrm{n}\Omega.}}
\end{equation}

\remove{Achieving a $R_s$ of this magnitude requires effective suppression of the residual-loss mechanisms discussed in the preceding sections, including trapped magnetic flux, seam losses, surface roughness, TLS-related losses, sub-gap states, and other fabrication-induced imperfections. Whether a $R_s$ of this magnitude can be achieved under the operating conditions considered in this work remains uncertain, since the present resonators operate at millikelvin temperatures, under low incident microwave powers, and without an accelerating field. In particular, the contributions of TLS losses and sub-gap states under these operating conditions remain insufficiently understood and must ultimately be determined experimentally.}

\subsection[\appendixname~\thesubsection]{Surface-Resistance Estimate for Additively Manufactured Niobium}
\label{app:Rs_3D_Nb_SRF}

The AdM resonator considered in this work is intended for fabrication from LPBF-fabricated Nb--47Ti (wt.\%). The superconducting performance of this material was recently investigated by Ersoz \textit{et al.}~\cite{Ersoz2026}. Although superconducting operation was successfully demonstrated, the measured microwave $R_s$ remained dominated by residual experimental loss mechanisms, predominantly trapped magnetic flux. Consequently, the intrinsic microwave performance achievable following active magnetic-field compensation, elimination of seam losses, and optimisation of the fabrication and surface-preparation procedures remains unknown. At present, no reliable estimate of the achievable microwave $R_s$ of LPBF-fabricated Nb--47Ti is therefore available.

In the absence of such an estimate, we adopt the lowest experimentally demonstrated microwave $R_s$ reported for any AdM niobium SRF cavity. This value was achieved using electron-beam melting (EBM), with Frigola \textit{et al.}~\cite{Frigola2015} reporting
\begin{equation}
    R_s = 170 \pm 10~\mathrm{n}\Omega,
\end{equation}
following buffered chemical polishing of a $3.9~\mathrm{GHz}$ cavity operated at approximately $2~\mathrm{K}$ and accelerating gradients of approximately $1~\mathrm{MV\,m^{-1}}$~\cite{Frigola2015}. The authors noted that the measured $R_s$ remained limited by fabrication-related factors including surface roughness, reduced density, microstructural imperfections, and material purity, and that further improvements in fabrication and post-processing would be required to achieve higher-performance SRF cavities.

Whether a surface resistance of this magnitude can be achieved under the operating conditions considered in this work remains uncertain, since the present resonators operate in a substantially different regime, namely at millikelvin temperatures, under low incident microwave powers, and without an accelerating field. Consequently, this value should be regarded as an optimistic estimate based on the best experimentally demonstrated performance of an AdM niobium SRF cavity rather than an intrinsic lower limit. Continued improvements in AdM processes, feedstock quality, oxygen control, relative density, surface finish, and post-processing may substantially reduce the achievable $R_s$.

For the purposes of calculating $\hat Q_0^{\mathrm{opt}}$, the $\hat R_s$ adopted for AdM cavities is therefore
\begin{equation}
    \boxed{\hat{R}_s = 170~\mathrm{n}\Omega.}
\end{equation}

\subsection[\appendixname~\thesubsection]{Summary}

Under the millikelvin-temperature and low-power operating conditions considered in this work, $R_{\mathrm{BCS}}$ is expected to be negligible, such that the total $R_s$ is dominated by residual-loss mechanisms. Assuming effective suppression of trapped magnetic flux, elimination of seam losses through fully welded cavity structures, and optimised cavity fabrication and surface preparation, no strong intrinsic frequency dependence of $R_s$ is expected over the $1$-$20~\mathrm{GHz}$ frequency range considered.

For the purposes of calculating $\hat Q_0^{\mathrm{opt}}$ for use in the cavity optimisation algorithm, the following benchmark values are adopted:
\begin{align}
    \hat R_s &= 0.63~\mathrm{n}\Omega, && \text{SM bulk niobium,} \\
    \hat R_s &= 170~\mathrm{n}\Omega, && \text{AdM niobium.}
\end{align}
These values are used solely to estimate $F(\mathbf{x})$ from the FEM eigenmode solutions, thereby enabling comparison of the candidate cavity resonators. They should not be interpreted as expected operating values. The $R_s$ ultimately achieved by each fabricated cavity will depend on its material quality, surface condition, fabrication and post-processing route, and experimental operating conditions, and must therefore be determined experimentally.

\section[\appendixname~\thesection]{Sensitivity to Pump AM-Noise}
\label{app:SpumpAM}
\begin{figure}[!t]
    \centering
    \includegraphics[width=0.3\textwidth]{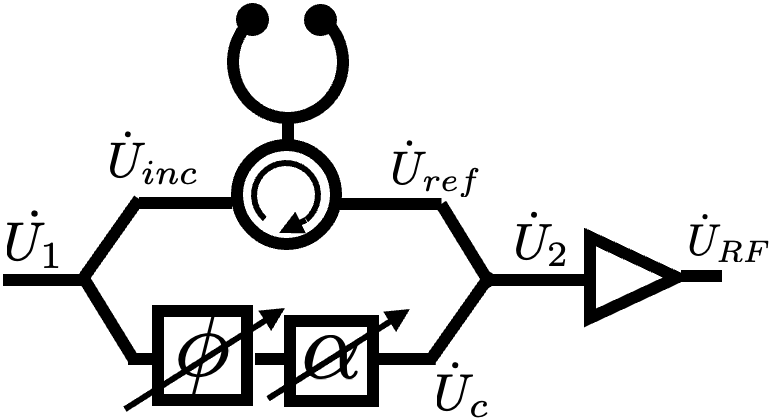}
    \caption{\label{fig:interferometerModel} Model of the interferometer showing definitions of the complex phasors.}
\end{figure}
To find $S_{AD}^p$ we assume the pump-AM noise is the only noise source. We can define the signal at the interferometer input as a complex phasor in the rotating frame of the pump frequency $\omega_0$ as
\begin{equation} \label{eq:U1}
    \dot{U}_1 = U_1 (1+m_p(\Omega))\cdot e^{j\phi_1}
\end{equation}
where $U_1$ is the amplitude, $\phi_1$ is the phase, and $m_{p}(\Omega)$ is some function of the angular Fourier frequency $\Omega$ that describes the pump AM-noise. The voltage incident on the resonator is simply 
\begin{equation} \label{eq:PumpSensUinc}
    \dot{U}_{inc} = (1/\sqrt{2}) \dot{U}_1
\end{equation}
and the reflected voltage is
\begin{equation}
    \dot{U}_{ref} = \Gamma \cdot \dot{U}_{inc}
\end{equation}
where $\Gamma$ is the reflection coefficient of a 1-port resonator given by
\begin{equation}
    \Gamma = \frac{\beta - 1 - j\xi}{\beta + 1 + j \xi}
\end{equation}
where $\beta$ is the coupling coefficient, $\xi = (\omega - \omega_0)/\Delta\omega_{0.5}$, where $\omega$ is the signal frequency, $\omega_0$ the resonant frequency and $\Delta\omega_{0.5}$ the intrinsic half-bandwidth of the resonator. The voltage at the output of the interferometer is
\begin{equation}
    \dot{U}_2 = \frac{1}{\sqrt{2}}\cdot \dot{U}_{ref} - \frac{1}{\sqrt{2}}\cdot\dot{U}_c
\end{equation}
and since we tune our compensation arm to cancel the carrier, i.e. 
\begin{equation}
\dot{U}_c = \dot{U}_{ref} (\omega_0) = \frac{\beta - 1}{\beta + 1} \cdot \dot{U}_{inc}
\end{equation}
the output of the interferometer is
\begin{equation}
\dot{U}_{2} = \frac{1}{\sqrt{2}} \cdot\frac{\beta - 1 - j\xi}{\beta + 1 + j\xi} \cdot \frac{1}{\sqrt{2}} \cdot\dot{U}_{1} - \frac{1}{\sqrt{2}}\cdot\frac{\beta - 1}{\beta + 1}\cdot\frac{1}{\sqrt{2}} \cdot\dot{U}_{1}.
\end{equation}
 which is indeed zero on resonance. At frequencies $\omega_0\pm\Omega$, substituting \eqref{eq:U1} setting $\phi_1 = 0$ and simplifying gives
\begin{equation}
    \dot{U}_2 (\omega_0 \pm \Omega) = \frac{-j\xi}{\beta + 1 + j\xi}\cdot\frac{2\beta}{\beta+1}\cdot\frac{U_1}{2}\cdot \frac{m_p(\Omega)}{2}
\end{equation}
for each sideband. Now taking the magnitude gives
\begin{equation}
    |\dot{U}_2 (\omega_0 \pm \Omega)| = \frac{\left(\frac{|\xi|}{\beta + 1}\right)}{\sqrt{1+\left(\frac{\xi}{\beta+1}\right)^2}}\cdot\frac{2\beta}{\beta+1}\cdot\frac{U_{inc}}{\sqrt{2}}\cdot \frac{m_{p}(\Omega)}{2}
\end{equation}
where we recall that $U_{inc} = U_1/\sqrt{2}$. Finally, multiplying by amplifier gain $K_{LNA}$, substituting $\xi = \Omega/\Delta\omega_{0.5}$, recalling that $\Delta\omega_{0.5}(1 + \beta) = \Delta\omega_{0.5}^L$ which is the loaded half-bandwidth, the square-root of power at the RF port of the mixer $\sqrt{P_{RF}}$ in each sideband at frequencies $\omega_0 \pm \Omega$ solely due to the pump AM-noise can be written as
\begin{equation}
  \begin{aligned}
    &\sqrt{P_{RF}^p (\omega_0 \pm \Omega)}\\ 
    &= \frac{|\Omega|/\Delta\omega_{0.5}^L}{\sqrt{1 + (\Omega/\Delta\omega_{0.5}^L)^2}}\cdot\frac{2\beta}{\beta+1}\cdot\sqrt{\frac{P_{inc} \cdot K_{LNA}}{2}} \cdot \frac{m_p(\Omega)}{2}
  \end{aligned}
\end{equation}
where $P_{inc}$ is the power incident on the resonator. In the AM-sensitive regime the mixer voltage $u_{mix} = \chi \sqrt{P_{RF}}$, so we obtain the expression for the sensitivity to pump AM-noise to be
\begin{equation}
  \begin{aligned}
    S_{AM}^p &= \frac{\delta u_{mix}}{\delta m_p} \\
    &= \chi\sqrt{\frac{P_{inc} \cdot K_{LNA}}{2}}\cdot\frac{|\Omega|/\Delta\omega_{0.5}^L}{\sqrt{1 + (\Omega/\Delta\omega_{0.5}^L)^2}}\cdot\frac{2\beta}{\beta+1}
  \end{aligned}
\end{equation}
where we have acquired an additional multiplication by 2 as both upper and lower sidebands add coherently when down-converted. Since axions produce modulation at frequency $\omega_a$ offset from the carrier we naturally set $\Omega = \omega_a$ giving~\eqref{eq:Spump} in the main text.

\section[\appendixname~\thesection]{Mixer Noise Floor Measurement} \label{app:mixer}
\begin{figure}[!t]
  \centering
  \includegraphics[width=0.49\textwidth]{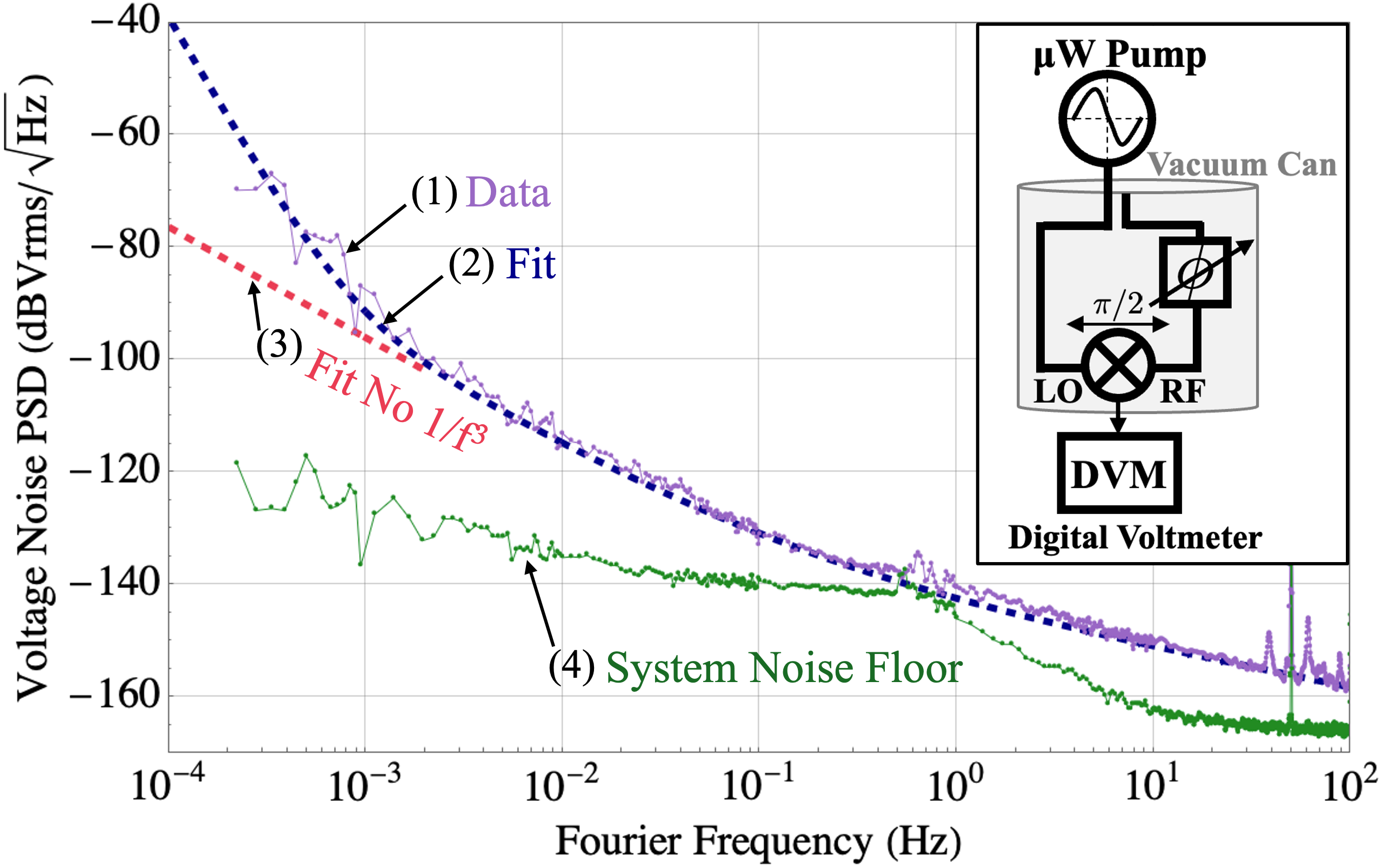}
  \caption{Mixer Noise Floor Measurement. \textbf{(a)} Schematic to measure mixer noise floor at low Fourier frequencies. \textbf{(b)} Trace 1 shows measured mixer noise at 10 GHz with -1.5 dBm RF power, and trace 4 shows the measurement system noise floor. Trace 2 shows a polynomial fit to the mixer noise, and trace 3 shows where this fit deviates from trace 2 when the $1/f^3$ term is removed.}
  \label{fig:mixer}
\end{figure}

To estimate the electronics noise floor we used a ML10220 double balanced mixer in a compact phase bridge with no device under test and tuned to be AM-insensitive. The whole phase bridge was then placed into a vacuum chamber and tested under a pressure of around 5 mbar to passively filter temperature and reduce the corner frequency for noise $\propto 1/f^3$. Below 0.5 Hz the mixer output voltage was sampled for 10 hours at a rate of 1 Hz and recorded with a Keysight 34465A digital voltmeter. Above 0.5 Hz the voltage was measured using a HP89410A vector signal analyzer and SR560 low-noise preamplifier. Both measurements had strong agreement at 0.5 Hz. The resultant mixer trace 1 was measured at -1.5 dBm RF power, and trace 4 shows the measurement system noise floor. 

A polynomial fit of the form \remove{$b_3/f^3 + b_1/f + b_{0.5}/f^{0.5}$}\review{$b_3/f^3 + b_1/f + b_{0.3}/f^{0.35}$} is shown in trace 2. The corner frequency between $1/f^{0.35}$ and $1/f$ noise is around 1 Hz, and between $1/f$ and $1/f^3$ noise is around 1 mHz. To estimate the ultimate mixer noise used in Section~\ref{sec:ProjSensitivity}, we assume that with addition of active temperature stabilisation the corner frequency for onset of $1/f^3$ noise can be further reduced to below 0.1 mHz such that the mixer noise follows trace 3 in this region. 

To estimate the interferometer noise, a noise floor measurement with a 32 dB gain Low-Noise-Factory amplifier at the mixer RF port was performed above 0.1 Hz with a 50 $\Omega$ terminator at the amplifier input. Though not shown in Fig.~\ref{fig:mixer}, this trace closely followed mixer noise below 10 Hz. Above this region the amplifier added a white noise level of -149 $\mathrm{dBVrms}/\sqrt{\mathrm{Hz}}$. Hence, at frequencies below 10 Hz, we infer that the interferometer noise would be the same as the mixer noise, and above 10 Hz we add the expected amplifier white noise to arrive at the inferred level for the interferometer electronics noise floor used for trace 1 of figure~\ref{fig:vNF}.

% \bibliography{your_external_BibTeX_file}
%\bibliography{References/UPLOAD_ANYON_refs.bib}

%=====================================
% References, variant B: internal bibliography
%=====================================

% % For the MDPI journals use author-date citation, please follow the formatting guidelines on http://www.mdpi.com/authors/references
% % To cite two works by the same author: \citeauthor{ref-journal-1a} (\citeyear{ref-journal-1a}, \citeyear{ref-journal-1b}). This produces: Whittaker (1967, 1975)
% % To cite two works by the same author with specific pages: \citeauthor{ref-journal-3a} (\citeyear{ref-journal-3a}, p. 328; \citeyear{ref-journal-3b}, p.475). This produces: Wong (1999, p. 328; 2000, p. 475)

% %%%%%%%%%%%%%%%%%%%%%%%%%%%%%%%%%%%%%%%%%%
% %% for journal Sci
% %\reviewreports{\\
% %Reviewer 1 comments and authors’ response\\
%Reviewer 2 comments and authors’ response\\
%Reviewer 3 comments and authors’ response
%}
%%%%%%%%%%%%%%%%%%%%%%%%%%%%%%%%%%%%%%%%%%

\end{document}